\documentclass[a4paper,11pt]{article}
\usepackage{verbatim}

\usepackage{dsfont}
\usepackage{sidecap}
\sidecaptionvpos{figure}{t}
\usepackage{subfigure}
\usepackage{physics}
\usepackage[utf8]{inputenc}
\usepackage[english]{babel}
\usepackage{empheq}
\usepackage{amssymb,amsmath}
\usepackage{graphicx, xcolor, varwidth}
\usepackage{setspace}
\usepackage[permil]{overpic}
\usepackage{cite}
\usepackage{graphicx}% see geometry.pdf on how to lay out the page. There's lots.
\usepackage{color}
\usepackage{braket}
\usepackage{simplewick}
\usepackage{jheppub}
\usepackage[T1]{fontenc}
\usepackage[normalem]{ulem}
\usepackage{mathrsfs}
\usepackage{amsbsy}

\usepackage[bottom]{footmisc}

\usepackage{simpler-wick}

\usepackage{feynmp}
\colorlet{darkblue}{blue!70!black}
\colorlet{darkgreen}{green!50!black}

\numberwithin{equation}{section}

\newcommand{\be}{\begin{equation}}
\newcommand{\ee}{\end{equation}}
\newcommand{\bea}{\begin{eqnarray}}
\newcommand{\eea}{\end{eqnarray}}
\newcommand{\bear}{\begin{eqnarray}}
\newcommand{\eear}{\end{eqnarray}}  
\newcommand{\beas}{\begin{eqnarray*}}
	
	\newcommand{\eeas}{\end{eqnarray*}}
\newcommand{\ba}{\begin{array}}
	\newcommand{\ea}{\end{array}}

\def\ba#1\ea{\begin{align}#1\end{align}}

\newcommand{\mW}{\mathcal{W}}

\renewcommand{\tr}{\operatorname{Tr}}
\newcommand{\pd}[2][1]{\ifnum#1=1 \frac{\partial}{\partial {#2}} \else
	\frac{\partial^#1}{\partial {#2}^{#1}}\fi}
\newcommand{\dpd}[2][1]{\ifnum#1=1 \dfrac{\partial}{\partial {#2}} \else
	\frac{\partial^#1}{\partial {#2}^{#1}}\fi}
\newcommand{\td}[2][1]{\ifnum#1=1 \frac{d}{d{#2}} \else
	\frac{d^#1}{d{#2}^{#1}}\fi}

\newcommand{\nbox}{{\,\lower0.9pt\vbox{\hrule \hbox{\vrule height 0.2 cm \hskip 0.19 cm \vrule height 0.2 cm}\hrule}\,}}

\newcommand{\mA}{\mathcal{A}}
\newcommand{\mH}{\mathcal{H}}
\newcommand{\mO}{\mathcal{O}}



		\title{	{ Seeing The Entanglement Wedge }}

			\author[a]{Adam Levine,}
			\author[b, c]{Arvin Shahbazi-Moghaddam,}
			\author[c]{Ronak M Soni}

		\affiliation[a]{Institute for Advanced Study, Princeton, NJ 08540, USA}
		\affiliation[b]{Center for Theoretical Physics and Department of Physics, University of California, Berkeley, CA 94720}
		\affiliation[c]{Stanford Institute for Theoretical Physics, 382 Via Pueblo, Stanford CA 94305}

		\emailAdd{arlevine@ias.edu}
		\emailAdd{arvinshm@stanford.edu}
		\emailAdd{ronakms@stanford.edu}
			
		\abstract{

	    We study the problem of revealing the entanglement wedge using simple operations. We ask what operation a semiclassical observer can do to bring the entanglement wedge into causal contact with the boundary, via backreaction.
	    
		In a generic perturbative class of states, we propose a unitary operation in the causal wedge whose backreaction brings all of the previously causally inaccessible `peninsula' into causal contact with the boundary.
		This class of cases includes entanglement wedges associated to boundary sub-regions that are unions of disjoint spherical caps, and the protocol works to first order in the size of the peninsula.
		The unitary is closely related to the so-called Connes Cocycle flow, which is a unitary that is both well-defined in QFT and localised to a sub-region. Our construction requires a generalization of the work by Ceyhan \& Faulkner to regions which are unions of disconnected spherical caps. We discuss this generalization in the Appendix. We argue that this cocycle should be thought of as naturally generalizing the non-local coupling introduced in the work of Gao, Jafferis \& Wall.
		
		}

\addtocounter{page}{1}

\begin{document}
	\maketitle 
	\parskip=10pt

\section{Introduction}
Recent developments in the study of AdS/CFT have revealed how a boundary CFT region $A$ encodes a large region of the bulk known as its entanglement wedge $\mathcal{W}_{E}[A]$ \cite{Headrick:2014cta,Jafferis:2015del,Dong:2016eik,Bao:2016skw,Faulkner:2017vdd,Cotler:2017erl,Chen:2019gbt,Kang:2018xqy,Gesteau:2020rtg}. Given an asymptotically $AdS$ spacetime $\mathcal{M}$, we can ascribe an effective field theory Hilbert space to the bulk degrees of freedom on $\mathcal{M}$. Entanglement wedge reconstruction embeds the effective field theory degrees of freedom in $\mathcal{W}_E[A]$ into the CFT degrees of freedom in $A$.

Entanglement wedge reconstruction does not seem constrained by the bulk causal structure; the causal wedge $\mathcal{W}_{C}[A]$ defined as the intersection of the past and the future of $D(A)$ (the domain of dependence of $A$) is generically a proper subset of $\mathcal{W}_{E}[A]$ \cite{maximin}. Let $\mathcal{W}_{\overline{C}}[A]$ denote the bulk wedge spacelike to $\mathcal{W}_{C}[A]$. It is easy to see that gravity is essential for the encoding of $\mathcal{W}_{E}[A] \cap \mathcal{W}_{\overline{C}}[A]$ to be possible. If we freeze gravity while holding $\mathcal{M}$ fixed, the bulk Hilbert space reduces to that of QFT on a curved background $\mathcal{M}$ and therefore all bulk operators in $\mathcal{W}_{C}[A]$ would exactly commute with CFT operators in $D(A)$, rendering bulk reconstruction impossible in $\mathcal{W}_{E}[A] \cap \mathcal{W}_{\overline{C}}[A]$. Indeed, it is believed that due to gravity, operators in $\mathcal{W}_{\overline{C}}[A]$ gain small commutators with simple CFT operators in $D(A)$ \cite{Harlow:2016vwg,Almheiri:2014lwa,HaPPY}. The current bulk reconstruction procedures, intuitively speaking, combine simple boundary operators with more complicated ones, like the modular Hamiltonian, so as to create large commutators with bulk operators everywhere in $\mathcal{W}_{E}[A]$.

An over-arching motivation behind this work is to seek a Lorentzian bulk interpretation of entanglement wedge reconstruction. Such a reconstruction is well-understood for operators supported within $\mathcal{W}_{C}[A]$. There are explicit expressions, often referred to as HKLL, for boundary operators equivalent to local bulk operators \cite{HKLL,Papadodimas:2012aq,Morrison:2014jha} which only utilize bulk dynamics.\footnote{These papers do so in specific symmetric backgrounds, but we expect that reconstruction of operators in $\mathcal{W}_{C}[A]$ continues to hold in any asymptotically AdS geometry, though explicit formulae may be hard to derive. We believe the time-like tube theorem \cite{borcherstt,arakitt}, see also \cite{rehren1}, provides hints for this. We thank Victor Gorbenko, Karl-Henning Rehren and Edward Witten for discussions on this point.} A more non-trivial question is whether there exists such a bulk interpretation of boundary reconstruction for operators in $\mathcal{W}_{E}[A] \cap \mathcal{W}_{\overline{C}}[A]$. Such an interpretation needs to exclusively involve bulk dynamics and is therefore constrained by bulk causality. A natural way to impose this constraint is to only consider procedures that do not perturb $\mathcal{W}_{E}[A] \cap \mathcal{W}_{\overline{C}}[A] \subset \mathcal{M}$; such operations could in principle change the geometry elsewhere so that this region (or parts of it) could come into causal contact with $D(A)$ in the perturbed geometry.\footnote{Note that all such perturbed geometries contain a region diffeomorphic to $\mathcal{W}_{E}[A] \cap \mathcal{W}_{\overline{C}}[A]$ in the original geometry $\mathcal{M}$ by definition. Therefore, it makes sense to refer to this ``same'' region in the perturbed geometry, though it would no longer be $\mathcal{W}_{E}[A] \cap \mathcal{W}_{\overline{C}}[A]$ of the new geometry} It is therefore an interesting question to ask if this can be realized in general.\footnote{Similar questions were asked in \cite{Engelhardt:2018aa} that partly motivated this work. Further, the role of backreaction and semi-classical operations was also emphasised in \cite{SR1,SR2,SR3}.}

A first step towards a Lorentzian bulk interpretation of entanglement wedge reconstruction would be to use such semi-classical operations to bring parts of $\mathcal{W}_E [A] \cap \mathcal{W}_{\overline{C}} [A]$ into \emph{either} the causal past or causal future of $D(A)$. This would make explicit the non-commutativity of simple CFT operators in $D(A)$ with bulk operators in $\mathcal{W}_E [A] \cap \mathcal{W}_{\overline{C}} [A]$. When a bulk region is in the causal past/future of $D(A)$, we say that $D(A)$ can `see'/`influence' that bulk region.

A basic example of such a procedure is the following, as pointed out in \cite{AAL}.\footnote{A different procedure for reconstruction in AdS-Vaidya geometries was discussed in \cite{Roy:2015pga}.}
Consider an AdS-Vaidya geometry, that differs from a static AdS-Schwarzschild geometry only by the presence of a shock and let $A$ be the entire boundary.
Removing the shock is then a semi-classical operation whose back-reaction expands the causal wedge and makes it agree with the entanglement wedges of the boundary CFT. A second example was provided in the seminal work of Gao, Jafferis \& Wall \cite{GJW}. If we consider a two-sided black hole in the thermo-field double state, there is an entire region behind the horizon which is inaccessible to left and right observers individually. Coordinated action between these left and right observers, however, can send in a pulse of negative energy that moves the causal horizon so that the observers may access some of the interior. While this may not seem to be a direct example of seeing the entanglement wedge since the black hole interior of the thermo-field double is already in the future of the boundary, in Section \ref{sec:peninsula} we discuss a simple variant of this set-up, first described in \cite{AMM}, where backreaction can bring previously space-like separated regions into causal contact with the boundary.\footnote{A different variant where the same effect is achieved can be found in \cite{Kourkoulou:2017zaj,Almheiri:2018ijj}.}

\begin{figure}%
    \centering
{\includegraphics[width=.5\columnwidth]{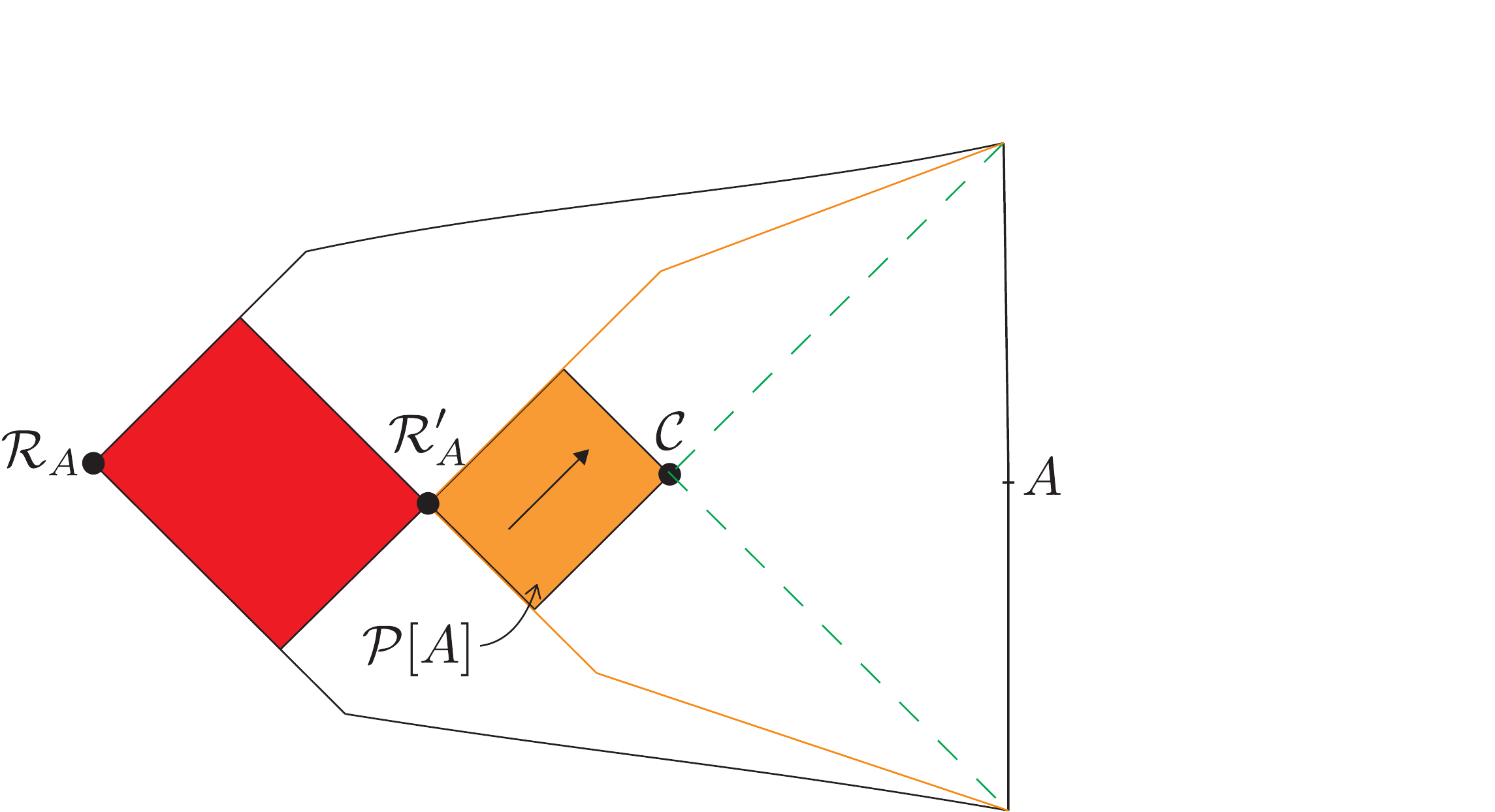} }%
    \qquad
{\includegraphics[width=.4\columnwidth]{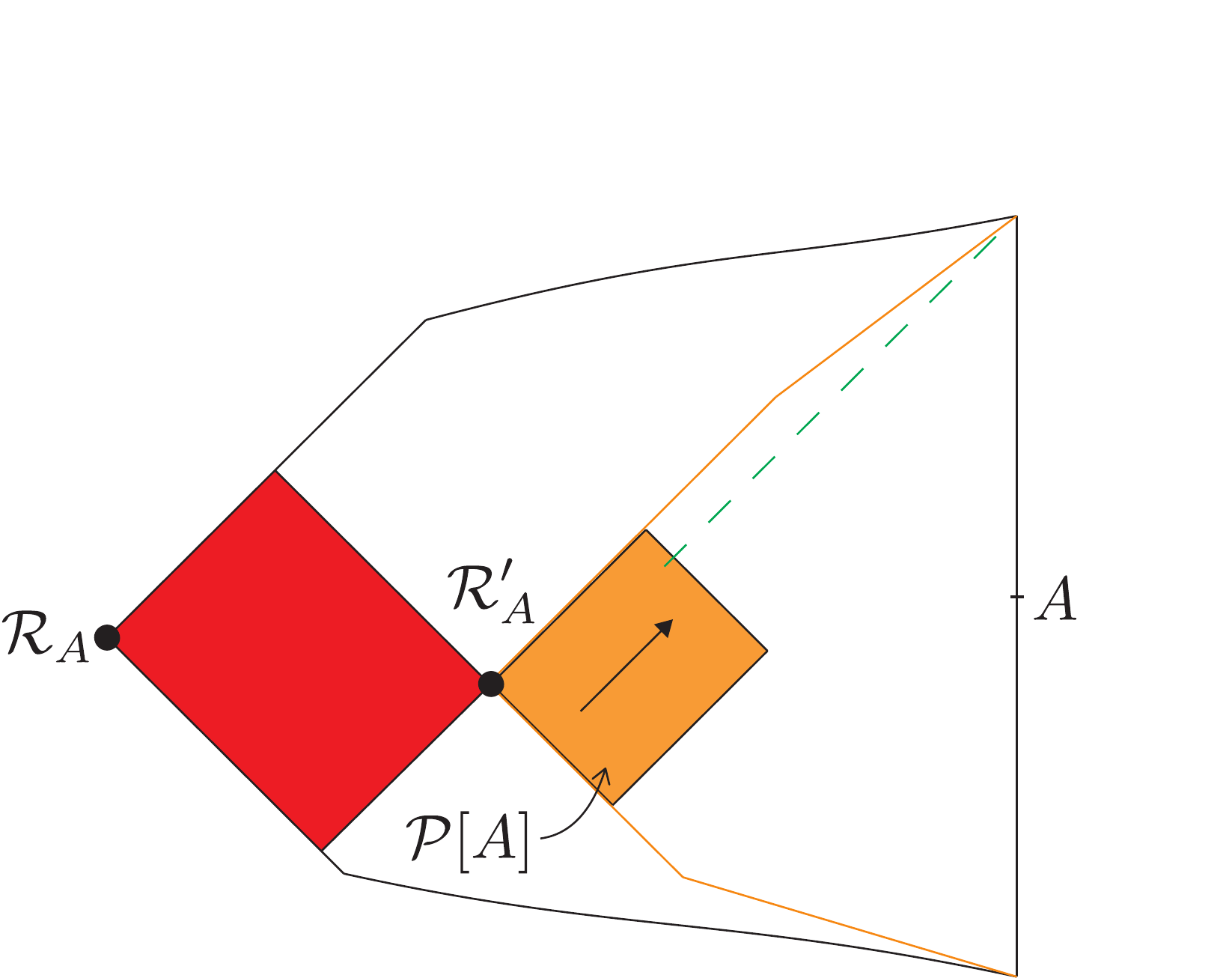} }%
    \caption{Left: given a boundary region $A$, the RT surface $\mathcal{R}_{A}$ and the corresponding entanglement wedge $\mathcal{W}_{E}[A]$(with black boundary), the outermost extremal surface $\mathcal{R}'_{A}$ and the corresponding outermost wedge $\mathcal{W}_{O}[A]$ (with orange boundary) is shown. The green dotted lines mark the causal horizons and therefore the boundaries of $\mathcal{W}_{C}[A]$. The intersection of the horizons defines the causal surface $\mathcal{C}$. Right: We consider bulk perturbations that don't change the state in $\mathcal{W}_{\overline{C}}[A]$, but change the geometry so as to bring parts of $\mathcal{W}_{E}[A] \cap \mathcal{W}_{\overline{C}}[A]$ into the causal past of $D(A)$ as shown. A signal (marked by the black straight arrow) can then reach $D(A)$. The causal past of $D(A)$ cannot be extended beyond $\mathcal{R}'_{A}$ under this class of operations. The peninsula $\mathcal{P}[A]$, marked in orange, is therefore the largest region that can be placed in the causal past (or future) of $D(A)$ using such operations, while the red region can never be accessed this way.}%
    \label{fig:main}%
\end{figure}

    Before going further, we need to explain an important limitation to this procedure. Let $\mathcal{R}_{A}$ be a quantum extremal surface homologous to $A$ and let $\mathcal{W}_{\mathcal{R}}[A]$ be the bulk domain of dependence of a homology slice of $\mathcal{R}_{A}$.\footnote{When $\mathcal{R}_{A}$ is also the minimum generalized entropy quantum extremal surface, then $\mathcal{W}_{\mathcal{R}}[A]= \mathcal{W}_{E}[A]$} Assuming the quantum focusing conjecture \cite{Bousso:2015mna}, it was shown that $\mathcal{W}_{C}[A] \subseteq \mathcal{W}_{\mathcal{R}}[A]$ \cite{maximin, Engelhardt:2014gca}.\footnote{In fact, there is a more restrictive constraint on the causal past and future of $D(A)$ posed by the quantum marginal surfaces \cite{Engelhardt:2013tra, Akers:2019aa, Bousso:2020ab}, i.e. surfaces with vanishing quantum expansion only along one null direction. However, for the main result of this paper, which pertain to certain perturbative setups, the restriction caused by the outermost quantum extremal surfaces is equivalent to that of the outermost quantum marginally trapped surfaces. Therefore, we will not elaborate on this stronger restriction here.} The outermost (closest to the boundary) such $\mathcal{R}$ is especially important since it has the smallest wedge and therefore poses the most stringent restriction on extending the causal past or future of $D(A)$\footnote{The natural definition is if $\mathcal{W}_{\mathcal{R}}[A]$ doesn't contain any portion of a quantum extremal surface homologous to A. In general, one would have to show such an outermost wedge exists. But in several simple cases including the special case considered in this paper such outermost extremal wedges exist}. We will henceforth call it the outermost extremal wedge and denote it by $\mathcal{W}_{O}[A]$ (See Fig. \ref{fig:main}). In Section \ref{sec:discussion}, we come back to this restriction on our ability to expand the causal past or future of $D(A)$ into regions beyond $\mathcal{W}_{O}[A]$ and mention a possible connection with a recent proposal on restricted complexity in AdS/CFT\footnote{We thank Geoff Penington for pointing this out.} \cite{PL}.

    In light of this restriction, one needs to refine the question posed earlier. Defining the ``peninsula'', $\mathcal{P}[A] = \mathcal{W}_{O}[A] \cap \mathcal{W}_{\overline{C}}[A]$, we can ask to what extent causal operations (which by our definition do not affect $\mathcal{P}[A]$ directly) can bring $\mathcal{P}[A]$ into the past or the future of $D(A)$ in the perturbed geometry. In the classical regime, the importance of wedges similar to $\mathcal{W}_{O}[A]$ was highlighted in \cite{Engelhardt:2018aa} where the authors put forth the idea that simple boundary sources might bring the wedge into causal contact with the boundary.\footnote{More precisely, the authors considered certain marginally trapped surfaces and their corresponding outer wedges}

In this paper, we investigate this question in a very special set up. We consider bulk semiclassical states and pick $A$ such that $\mathcal{W}_{C}[A]$ and $\mathcal{W}_{O}[A]$ agree classically, but differ due to bulk quantum effects. More specifically, we consider states for which $\mathcal{P}[A]$ scales like the Planck length multiplied by a large number $c$; a simple way to achieve this is to consider a bulk theory that has a large number $c$ of light fields. We then show that specific instances of bulk unitary operations known as Connes cocyle (CC) flow have interesting properties that upon backreaction could bring $\mathcal{P}[A]$ in the causal past or future of $D(A)$.\footnote{This should be contrasted with the bulk dual of a boundary CFT cocycle flow studied in \cite{Bousso:2020aa}. We will return to boundary cocycle flow in Section \ref{sec:bd}.} We will demonstrate this explicitly in a more limited setting where the background geometry is exactly pure $AdS$ and the boundary region, A, is a union of disconnected spheres. Inspired by the properties of this unitary transformation, we show how analogous properties in general spacetimes perturbatively different from $AdS$ are sufficient to bring $\mathcal{P}[A]$ to the causal past (or future) of $D(A)$. We then conjecture the existence of such a unitary.

In pure AdS, the non-empty $\mathcal{P}[A]$ is present because of bulk mutual information between the different bulk regions. As we will discuss in detail in Sec. \ref{sec:cc}, the CC flow generates negative energy shocks at the causal surface which are present precisely due to the mutual information and are of the exact right magnitude to bring $\mathcal{P}[A]$ to causal contact with the boundary. More broadly, when the geometry is perturbatively close to AdS, the CC flow acts also as an AdS-Rindler boost in the causal wedge which reduces the time delay by diluting the infalling excitations on the horizon. This will in turn bring $\mathcal{P}[A]$ in causal contact with the boundary.

The Connes-cocyle-like operator we shall define is qualitatively different from Petz-map-like reconstructions of the bulk; the Petz map is best thought of in terms of \emph{Euclidean, boundary} modular evolution \cite{Penington:2019kki} and could therefore be exponentially complex --- $\mO\left(e^{G_N^{-1}}\right)$ --- when written in terms of simple unitaries on the boundary \cite{PL}.
The operator we define will be defined in terms of \emph{causal, bulk} modular evolution, and so is correspondingly simple.
In more detail: it will be defined entirely in the bulk effective field theory, which means that it exists in the limit $G_N \to 0$ as an operator in the bulk QFT, and therefore has a complexity bounded by $\mO(G_N^0)$ when written in terms of elementary bulk local operators on a Cauchy slice of the causal wedge.
Finally, elementary local operators in a Cauchy slice of the causal wedge can be written as elementary local operators in a Cauchy slice of the boundary causal diamond using the HKLL prescription and so have a complexity that is $\mO(\text{poly} (G_N^{-1}))$ in terms of these operators.

We now summarize the rest of the paper and its main results.

\begin{itemize}

\item In Sec. \ref{sec:cc}, we discuss the CC unitary flow from a purely field theoretic point of view. We will summarize the main properties of the flowed states and derive the existence of certain stress tensor shocks. These results are a generalization of the derivations in \cite{CF} to the case of multiple regions. In Appendix \ref{app:aqft}, we re-derive these results using algebraic QFT techniques applied to general von Neumann algebras which relaxes the assumption used in Sec. \ref{sec:cc} that the Hilbert space of QFT factorizes. In Appendix \ref{app:kink}, we derive the stress tensor shocks using a holographic technique following (and generalizing) the results of \cite{Bousso:2020aa}.

\item In Section \ref{sec:peninsula}, we examine in detail an example where the background is exactly pure $AdS$ and show how cocycle flow induces stress tensor shocks of just the right magnitude to expand the causal past or the future of $D(A)$, revealing $\mathcal{P}[A]$. Specifically, we consider the setup in \cite{AMM, AMMZ} of JT gravity in thermal equilibrium with the bath and pick $A$ to be a disjoint union of two regions. Then, $\mathcal{P}[A]$ will be perturbatively small and the negative energy shocks bring $\mathcal{P}$ into causal contact with the boundary in a way analogous to the Gao-Jafferis-Wall protocol \cite{GJW}. In Appendix \ref{app:jt} we review various aspects of JT gravity.

\item In Section \ref{sec:gen}, we will take lessons from the examples of Sec. \ref{sec:peninsula} and conjecture the existence of an analogous unitary transformation in more general cases, including those in which the background perturbatively differs from pure $AdS$. Our conjecture is motivated by the fact that the properties exactly mirror that of the cocycle flow in the fixed $AdS$ background. We then show that this conjectured transformation can extend the causal past or the future, bringing $\mathcal{P}[A]$ in causal contact with $D(A)$. We then propose a possible construction of such a unitary, highlighting the main hurdles in the way of proving that this realizes our conjecture.

\item In Section \ref{sec:bd}, we explore how our picture dovetails with previous discussions of bulk reconstruction using modular flow \cite{Faulkner:2017vdd,Cotler:2017erl,Chen:2019gbt}. In particular, we show that a boundary version of the CC flow operator, which is related to the discussions in \cite{Faulkner:2017vdd}, renders exactly the same set of operators visible to the boundary as in Sec. \ref{sec:peninsula}, the AdS-Vaidya example mentioned above and Sec. \ref{sec:gen}. This relation is only true when the extremal surface close to the causal surface is in fact the true minimal surface, i.e. when the boundary relative entropy between $\ket{\psi}$ and the split state does not scale with $N^2$; this is unlike the previous sections which work even when the outermost extremal surface is non-minimal.

\item In Sec. \ref{sec:discussion}, we briefly discuss a connection to the recent story of \cite{PL} and the failure to see beyond the outermost extremal wedge using the causal operations described here. Furthermore, we discuss whether our procedure generalizes to even more examples, such as entanglement wedges associated to arbitrary boundary regions. We also discuss to what extent bringing $\mathcal{P}[A]$ into the causal past or future of $D(A)$ but not both simultaneously is related to the notion of bulk reconstruction.

\end{itemize}

\section{Summary of the Connes cocycle flow} \label{sec:cc}

In this section we describe a class of unitaries in QFT known as the Connes' cocycle flow \cite{cc-connes,cc-araki,Bousso:2020aa,CF}. These unitaries applied to certain connected regions have recently been studied in \cite{CF}. In \cite{Bousso:2020yxi}, it was shown that the flow induces certain stress tensor shocks with magnitudes proportional to shape derivatives of the von Neumann entropy. Here, we will generalize this flow to include multiple regions. In subsection \ref{sub-summary}, we summarize some salient features of the flow, in particular the presence of analogous stress tensor shocks to the ones found in \cite{Bousso:2020yxi, CF}. In subsection \ref{sub-shocks}, we will give a derivation of the shocks, assuming some smoothness conditions on the relative entropy following \cite{CF} which we will derive in Appendix \ref{app:aqft}. Throughout this section, we assume the existence of a locally factorizable Hilbert space structure in QFT enabling us to discuss density matrices associated to QFT subregions. In Appendix \ref{app:aqft}, we will re-derive the results of this section in algebraic QFT language and for general von Neumann algebras, relaxing the factorization assumption.

\subsection{Summary of the flow and its properties}\label{sub-summary}

Consider QFT on some fixed $d+1$ dimensional background. Let $a$ be the union of $n$ disjoint, co-dimension one regions, $a=\cup_{r=1}^{n} a_r$, such that for each $r$ there exists some boost-like Killing field $\xi$ which preserves $D(a_r)$. For concreteness, we will from now on focus on fixed $AdS_{d+1}$ background and let each $D(a_r)$ be an AdS-Rindler wedge. But the analogous results below apply to any such $a$.

\begin{figure}
    \centering
    \includegraphics[width=0.4\columnwidth]{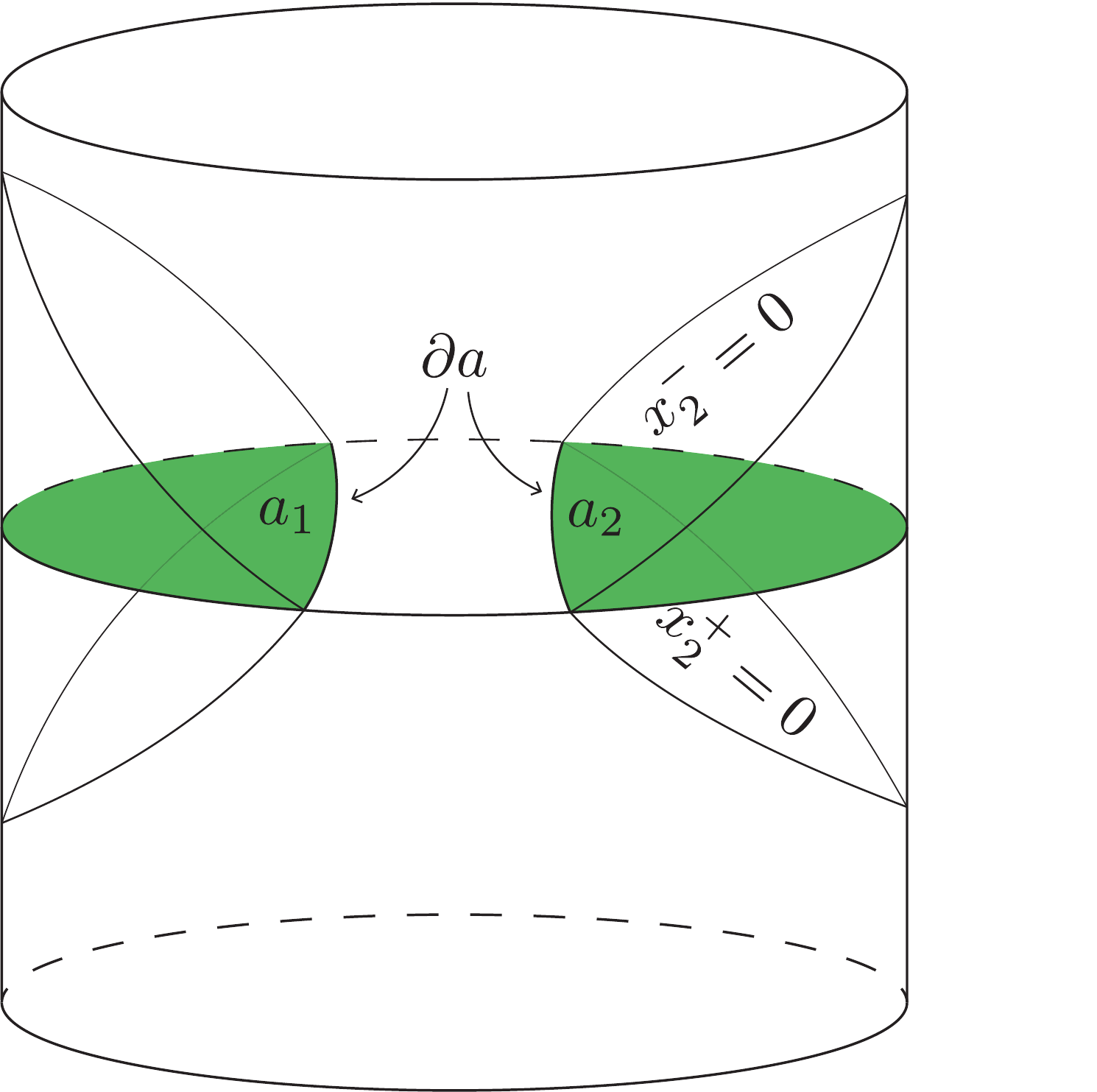}
    \caption{Region $a=a_{1}\cup a_{2}$ is shown in orange. Each $D(a_{r})$ is an AdS-Rindler wedge whose future and past boundary lies on a Killing horizon. For $D(a_{2})$, the horizons are marked by $x^-_2=0$ and $x^+_2=0$ respectively.}
    \label{fig:twoAdSRindlers}
\end{figure}

We can parametrize each $D(a_r)$ with coordinates $(x^+_r, x^-_r, y^i_r, z_r)$ with $x^+_r\geq 0$ and $x^-_r\leq 0$ null affine parameters on the future and past boundaries of $D(a_r)$ and such that $x^+_r=x^-_r=0$ marks $\partial a_r$ while $z_r$ and $y^i_r$ for $i=1,\cdots, d-2$ parametrize the transverse direction (along $\partial a_r$). Poincare coordinates are one such choice of coordinates for the AdS-Rindler wedge for $x^+_r\geq 0$ and $x^-_r\leq 0$ and the metric is
\begin{align}
    ds^2 = \frac{1}{z_r^2}\left( dz_r^2- dx^-_r dx^+_r + \sum_{i=1}^{d-2} (dy^{i}_r)^2 \right).
\end{align}
Here and later, we set the $AdS$ radius to $\ell = 1$.
Furthermore, in these coordinates, we have
\begin{align}
   \xi = x^+_r\partial_{x^+_r} - x^-_r\partial_{x^-_r}
\end{align}
and the isometry $\Phi^{\xi}_{s}$ generated by $\xi$ is
\begin{align}\label{eqn:boost}
    \Phi^{\xi}_{s}(x^+_r,x^-_r,y^i_r, z_r) = (x^+ e^{2\pi s} ,x^- e^{-2\pi s},y^i_r, z_r).
\end{align}

Now consider a pure state $\ket{\psi}$. Given some region $\Sigma$, we can trace out the complement region $\bar{\Sigma}$, and define the density matrix $\rho^{\psi}_{\Sigma} = \tr_{\bar{\Sigma}} \ket{\psi}\bra{\psi}$. In the special case of the vacuum state $\ket{\psi}=\ket{\Omega}$, we use the notation $\sigma_{\Sigma} = \tr_{\bar{\Sigma}} \ket{\Omega}\bra{\Omega}$. It is well known that for $a_r$, $\sigma^{is}_{a_r}$ generates the flow $\Phi^{\xi}_{s}$ on local operators in $D(a_{r})$ \cite{Bisognano:1975kp,Morrison:2014jha}:
\begin{align}
 \sigma_{a_{r}}^{is} \mathcal{O} (x) \sigma_{a_{r}}^{-is} = \mathcal{O}(\Phi^{\xi}_{s}(x))
\end{align}
where $\mathcal{O} (x)$ is a local operator in $D(a_{r})$.

The Connes cocycle unitary transformation can now be defined in the following way:
\begin{align}\label{eq-flowbasic}
\ket{\psi_s} := \otimes_{r=1}^{n} \sigma_{a_{r}}^{is} (\rho^{\psi}_{a})^{-is} \ket{\psi}
\end{align}
where $\rho_{a}$ is the reduced density matrix of $a$ in the state $\ket{\psi}$. The $n=1$ case of this unitary was recently studied in \cite{CF}. Let
\begin{align}\label{eqn:cocycle}
u_{s}(\psi,a) := \otimes_{r=1}^{n} \sigma_{a_{r}}^{is} (\rho^{\psi}_{a})^{-is}
\end{align}
from now on.

We will now state some important properties of this flow, delaying the justification for some of them to the next subsection and the appendix. Since this is a unitary transformation with support restricted to $D(a)$, the observables restricted to the complementary region $D(\bar{a})$ (the region spacelike to $D(a)$) are preserved under this transformation. On the other hand, the observables restricted to $a$ get transformed by the isometry flow $\Phi^{\xi}_{s}(\cdot)$ generated by the killing field $\xi$ in the following sense
\begin{align}\label{eq-boost-transform}
\bra{\psi_s} \mathcal{O}(x_{1}) \cdots \mathcal{O}(x_{N}) \ket{\psi_s} = \bra{\psi} \mathcal{O}(\Phi^{\xi}_{-s}(x_{1})) \cdots \mathcal{O}(\Phi^{\xi}_{-s}(x_{N})) \ket{\psi}
\end{align}
where $x_{1},\cdots, x_{N}$ are points in $D(a)$. This follows from the cyclicity of trace and the fact that $\sigma_{a_{r}}^{is}$ generate $\Phi^{\xi}_{s}(\cdot)$. Note that for non scalar operators an appropriate transformation needs to be applied to the corresponding tangent bundle as well. Simple transformation rules for observables that are not restricted to either $D(a)$ or $D(\bar{a})$ are not known in general.

\begin{figure}
    \centering
    \includegraphics[width=0.4\columnwidth]{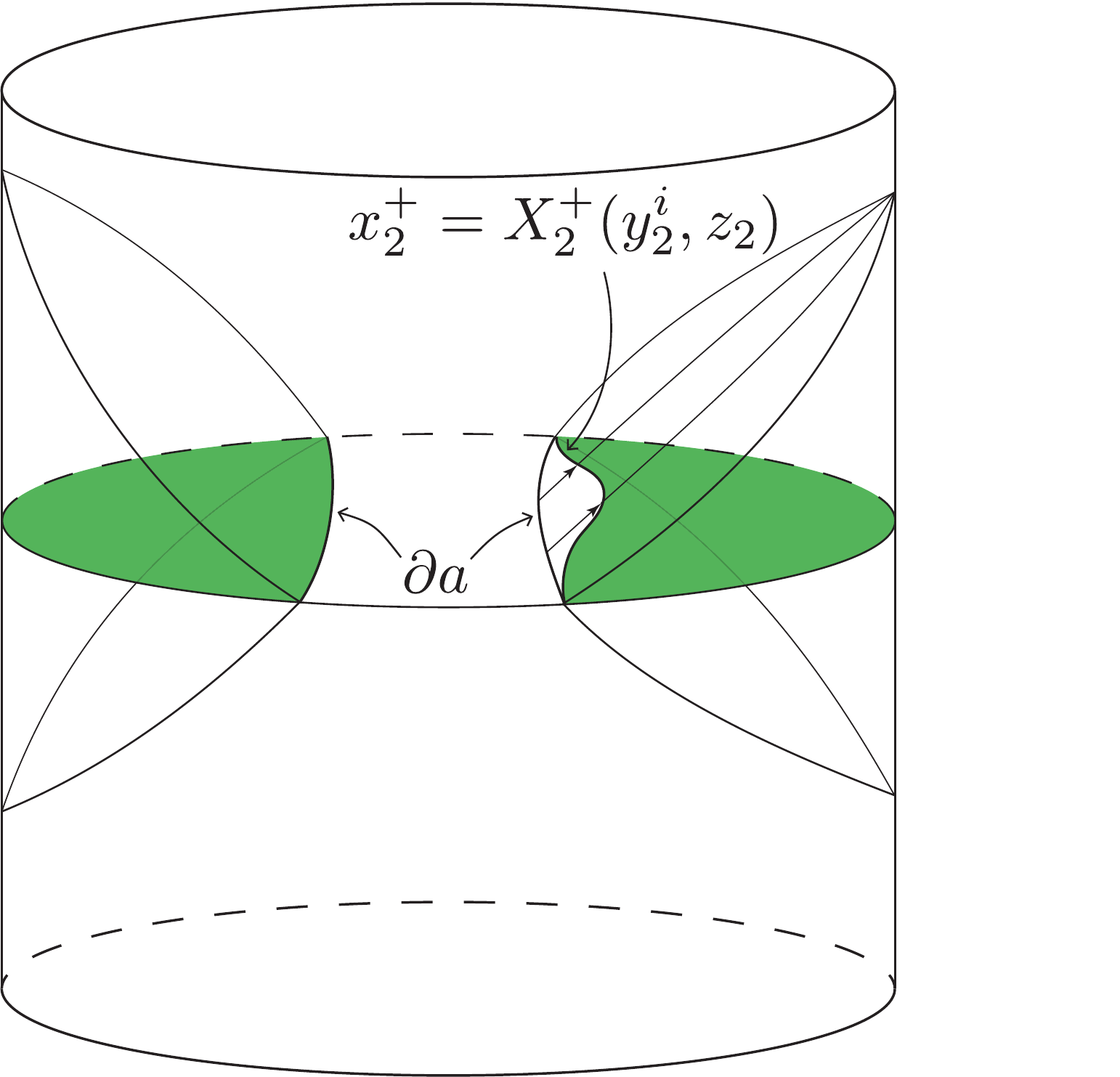}
    \caption{The green region marks $a(X^{+}_2)$ which is a deformation of $a$ by moving $\partial a_2$ in the null direction along $x^-_2=0$. The $\psi_s$ states have stress tensor shocks at $\partial a$ proportional to the derivative of the von Neumann entropy under such shape deformations.}
    \label{fig:twoAdSRindlerdeformed}
\end{figure}

Another salient feature of $\psi_s$ is the existence of stress energy tensor shocks at $\partial a$ for $s\neq0$. We will now merely state the magnitude of these shocks and provide the derivation in the next subsection. For simplicity, we also assume that the state $\ket{\psi}$ has a continuous stress tensor profile around $\partial a$. Then at $\partial a$ we have
\begin{align}\label{eqn:energyshocks}
&\bra{\psi_s}T_{++}(x^+_r,x^-_r=0,y^i_r, z_r) \ket{\psi_s} = \frac{1}{2\pi}(e^{-2 \pi s} - 1) \frac{1}{\sqrt{H(y^i_{r}, z_r)}} \left.\frac{\delta S(\rho^{\psi}_{a(X^+_r)})}{\delta X^+_r(y^i_r, z_r)} \right|_{X^{+}_r=0}\delta(x^+_r) + o(\delta)\\
&\bra{\psi_s}T_{--}(x^+_r=0,x^-_r, y^i_r, z_r) \ket{\psi_s} = \frac{1}{2\pi}(e^{2 \pi s}-1) \frac{1}{\sqrt{H(y^{i}_{r}, z_r)}} \left.\frac{\delta S(\rho^{\psi}_{a(X^-_r)})}{\delta X^-_r(y^i_r, z_r)}\right|_{X^{-}_r=0} \delta(x^-_r)+o(\delta)
\end{align}
where $H = z^{2-2d}$ denotes the determinant of the intrinsic metric of $\partial a$ and $o(\delta)$ denotes finite (non-distributional terms). We will now explain the RHS of Eqs. \eqref{eqn:energyshocks}. $S(\rho)$ denotes the von Neuman entropy of the density matrix $\rho$, and $a(X^{+}_r)$ denotes a deformation of the region $a$ whereby $a_{r}\in a$ is deformed along the $x_r^-$=0 horizon to a new region bounded by $(x_r^-=0, x_r^+=X_{r}^{+}(y^i_{r}, z_r))$(See Figure \ref{fig:twoAdSRindlerdeformed}). Though this does not uniquely fix $a(X^+_r)$, it uniquely fixes $D(a(X^+_r))$ which in turn fixes $S(\rho^{\psi}_{a(X^+_r)})$. $a(X^-_r)$ is analogously defined.

Von Neumann entropies of subregions notably have UV divergences which need to regulated by a some proper cut-off procedure. These divergences are proportional to local geometric terms on the boundary of the subregion. For Rindler-like regions, the shape derivative of these geometric terms vanishes rendering $\delta S / \delta X$ UV finite.

In the case of a conformal field theory (CFT), the regions $D(a_r)$ need only be preserved by a conformal Killing vector field for there to be properties similar to Eqs. \eqref{eq-boost-transform},\eqref{eqn:energyshocks} in $\psi_s$. This is because for a CFT, $\sigma_{a_{r}}$ will generate a local conformal Killing flow if $D(a_{r})$ is preserved by one. This enlarges the class of geometric regions allowed. For example, $a$ could be the union of disjoint ball-shaped regions in Minkowksi space. The transformation rule in Eq. \eqref{eq-boost-transform} then acquires an appropriate conformal rescaling of the operators as well.

\subsection{Deriving stress tensor shocks in $\psi_s$}\label{sub-shocks}
We will now explain how the stress tensor shocks in Eqs. \eqref{eqn:energyshocks} can be derived, following \cite{CF} closely. The explanation relies on a certain continuity condition of the shape derivative of the relative entropy that was originally shown in \cite{CF} when the number, $n$, of connected regions is $n=1$. In Appendix \ref{app:aqft}, we will extend the results to general $n$. Here we will state the result and derive Eqs. \eqref{eqn:energyshocks} using differentiability of the relative entropies. The relative entropy between $\rho$ and a reference state $\sigma$ is defined as
\begin{align}
    S_{\text{rel}}(\rho|\sigma) = \tr [\rho \log \rho] - \tr [\rho \log \sigma]
\end{align}
We can re-write the relative entropy in the following way:
\begin{align}
    S_{\text{rel}}(\rho|\sigma) = \tr [\rho \Delta H^{\sigma}] - \Delta S
\end{align}
where
\begin{align}
   \Delta H^{\sigma}  = -\log \sigma + \left( \tr [\sigma \log \sigma] \right) \mathds{1},\ \ \  
  \Delta S = S(\rho)-S(\sigma)
\end{align}
where $\Delta H^{\sigma}$ is the $\sigma$-subtracted modular Hamiltonian.

From now on, we take the reference state to be $\sigma_{a} = \otimes_{r=1}^{n} \sigma_{a_r}$. As we mentioned before, $\sigma_{a_r}^{is}$ generates $\Phi^{\xi}_{s}$. In particular, this means that the vacuum-subtracted modular Hamiltonian is given by the charge of the $\xi$ flow (the boost generator) in $D(a_r)$ \cite{Bisognano:1975kp}

\begin{align}\label{eqn:vacmodhamflat}
\Delta H_{a_{r}}^{\Omega} \equiv -\log \sigma_{a_{r}} + \bra{\Omega}\log \sigma_{a_{r}}\ket{\Omega}\mathds{1} = \sum_{r=1}^{n} 2\pi \int dz_r d^{d-2}y^i_r \sqrt{H(y^{i}_{r}, z_r)} \int_{0}^{\infty} dx^+_r x^+_r T_{++}(x^+_r,x^-_r=0,y^{i}_{r}, z_r).
\end{align}
It was shown in \cite{Casini:2017aa} that this local form persists for subregions $a_{r}(X^{+}_r)$ defined above. The modular Hamiltonian takes the form
\begin{align}\label{eqn:vacmodham}
\Delta H_{a_{r}(X^+_r)}^{\Omega} = \sum_{r=1}^{n} 2\pi \int d z_r d^{d-2}y^{i}_{r} \sqrt{H(y^{i}_{r}, z_r)} \int_{X^+_r(y^{i}_{r}, z_r)}^{\infty} dx^+_r (x^+_r -X^+_r(y^i_r, z_r)) T_{++}(x^+_r,x^-_r=0,y^i_r, z_r).
\end{align}

We can then take derivatives of this relative entropy with respect to variations of $X^+_r(y^i_r,z_r)$ in of $a_{r}$. In what follows, we will assume that all of the relative entropies are finite and that their derivatives with respect to $X^+_r(y^i_r, z_r)$ variations are everywhere continuous. 

The key ingredient for us will be the differentiability of relative entropies in the $\psi_s$ state. For now we assume this and show how it predicts the existence of the shocks \eqref{eqn:shocks}. Differentiability will be shown in Appendix \ref{app:sumrule}, generalizing the results of \cite{CF}. Differentiability of relative entropy between $\psi_{s}$ and $\otimes_{r=1}^{n} \sigma_{a_{r}}$  at the entangling surface means 
\begin{align}\label{eqn:differentiable}
\lim_{X^+_r \to 0^{-}}\frac{\delta}{\delta X^+_r(y^i_r)} S_{\text{rel}}(\rho^{\psi_{s}}_{a}|\otimes_{r'=1}^{n} \sigma_{a_{r'}})=\lim_{X^+_r \to 0^{+}} \frac{\delta}{\delta X^+_r(y^i_r)} S_{\text{rel}}(\rho^{\psi_{s}}_{a}|\otimes_{r'=1}^{n} \sigma_{a_{r'}}).
\end{align}
If we write the relative entropy in terms of density matrices as
\begin{align}
S_{\text{rel}}(\rho | \sigma) = \text{Tr}[\rho \log \rho] - \text{Tr}[\rho \log \sigma]
\end{align} 
it then follows from equations \eqref{eqn:vacmodham} and \eqref{eqn:differentiable} that
\begin{align}\label{eq-energyresult}
    \int_{0^-}^{\infty} dx^+_r \langle T_{++}(x^-_r=0, x^+_r, y^i_r, z_r)\rangle_{\psi_{s}} - \int_{0^+}^{\infty} &dx^+_r \langle T_{++}(x^-_r=0, x^+_r, y^i_r, z_r)\rangle_{\psi_{s}} =\nonumber\\ &\frac{1}{2\pi}(e^{-2\pi s}-1)\frac{1}{\sqrt{H(y^{i}_{r}, z_r)}}\left.\frac{\delta S(\rho^{\psi}_{a(X^+_r)})}{\delta X^+_r(y^i_r, z_r)} \right|_{X^{+}_r=0},
\end{align}
where we used that, due to equation \eqref{eq-boost-transform}, the von Neumann entropies in the $\psi_s$ state are simply related by a boost to the von Neumann entropies at $s=0$
  \begin{equation}
    \lim_{X^{+}_r \to 0^{+}} \frac{\delta S (\rho^{\psi_s}_{a(X^+_r)})}{\delta X^{+}_r(y^i_r, z_r)} = e^{-2\pi s} \lim_{X^{+}_r \to 0^{+}}\frac{\delta S (\rho^{\psi}_{a(X^+_r)})}{\delta X^{+}_r(y^i_r, z_r)}.
    \label{eqn:ee-boost}
  \end{equation}
From this and the analogous relation for deformations along the $x^{+}=0$ horizon we get the following stress tensor distribution at $\partial a$
\begin{align}\label{eqn:shocks}
\braket{T_{++}(x^+, x^-=0, y^{i}_{r}, z_r)}_{\psi_s} = \frac{1}{2\pi}(e^{-2\pi s} - 1)\frac{1}{\sqrt{H(y^{i}_{r}, z_r)}} \left.\frac{\delta S(\rho^{\psi}_{a(X^+_r)})}{\delta X^+_r(y^i_r, z_r)} \right|_{X^{+}_r=0} \delta(x^+)+ o(\delta) \\
\braket{T_{--}(x^+ =0, x^-, y^{i}_{r}, z_r)}_{\psi_s} = \frac{1}{2\pi}(e^{2\pi s}-1)\frac{1}{\sqrt{H(y^{i}_{r}, z_r)}} \left.\frac{\delta S(\rho^{\psi}_{a(X^-_r)})}{\delta X^-_r(y^i_r, z_r)} \right|_{X^{-}_r=0}  \delta(x^-) + o(\delta)
\end{align}

In Appendix \ref{app:kink}, we will re-derive these shocks using the holographic dual to the transformation \eqref{eq-flowbasic}, while also showing that the other components of the stress tensor do not get shocks. We will take this as evidence that in general the transformation \eqref{eq-flowbasic} will only result in the shocks in Eq. \eqref{eqn:energyshocks}. We will now turn to utilizing these shocks in the context of semi-classical gravity in order to see behind the horizon.

%%%%%%%%%%%%%%%%%%%%%%%%%%%%%%%%%%%%%%%%%%%%%%%%%%%%%%%%%%%%%%%%%%%%

\section{Seeing the Peninsula in JT Gravity} \label{sec:peninsula}

We now apply this construction to our first simple example. We illustrate an example in Jackiw-Teitelboim gravity which was first described in the work of \cite{AMM}. We review the relevant properties of JT gravity in Appendix \ref{app:jt}. We begin with a brief review of the set-up and then show how, using the Connes' cocycle with a particular sign of $s$ from the previous section, we are able to see up to within a planck distance of the quantum extremal surface.
While we explicitly focus only on the sign of $s$ that brings the peninsula in the past of the boundary, it will be evident that the other sign of $s$ would bring the peninsula to the future of the boundary.

\subsection{The Set-up}
Following the lead of \cite{AMM, AMMZ, AHMST}, consider a finite temperature black hole coupled to a flat space bath. The bulk field theory will be taken to be a conformal field theory with central charge $c$. We imagine that this black hole is dual to two entangled BCFTs, whose boundaries sit at some cut-off surface in the asymptotic region of the black hole. We will work in Kruskal-Szekeres coordinates which cover a whole Minkowski patch. The metric in the black hole region is 
\begin{align}
ds^2_{BH} = -\frac{4dw^+ dw^-}{(w^+w^-+1)^2},
\label{eqn:bh-metric}
\end{align}
where the dilaton takes the form
\begin{align}
\phi(w^+,w^-) = \frac{2\pi \phi_r}{\beta} \frac{1-w^+w^-}{1+w^+w^-}.
\label{eqn:bh-dilaton}
\end{align}
We take the BCFTs to lie along a time-like trajectory at $w^+w^-=-1+\epsilon$ with transparent boundary conditions connecting the black hole to flat space baths. The metric in the bath region is 
\begin{align}\label{eqn:bathmetric}
    ds^2_{bath}= -\frac{\beta^2}{4\pi^2} \frac{dw^+ dw^-}{w^+w^-}.
\end{align}
The overall conformal factor is present so that the bath metric is flat with respect to Rindler coordinates $w^{\pm} = \pm e^{\frac{2\pi y^{\pm}}{\beta}}$.

For now, we pick the state of the quantum fields to be in the vacuum on the manifold with metric $ds^2 = -dw^+ dw^-$. Assuming we have a conformal field theory, the transformation law for the stress tensor under Weyl re-scaling tells us that the chiral components of the stress tensor on the black hole background also vanish as they should. The stress energy $T_{\pm\pm}$ is constant, but non-zero, in the bath region.

We consider the region of the dual quantum theory given by $A = A_1 \cup A_2$, with $A_1$ the entire right BCFT and $A_2$ the region of the left BCFT that stretches from $-\infty$ to $w^- = - w^+ = W_1^-$. As usual, we use upper case letters for boundary/UV theory subregions and lower case letters for bulk/IR theory subregions.
The causal wedges of $A_{1},A_2$ are the right Rindler wedge $a_1$ and $a_2 = A_2$ respectively,\footnote{While $a_2$ and $A_2$ are identical regions in the $w$ plane, they differ as subregions of theories. $A_2$ is a region of the UV theory and contains all the CFT operators within that region, whereas $a_2$ is a region of the IR theory and only contains those operators that don't disturb the bulk geometry too much. We know from general considerations of quantum error correction that $a_2$ is isometrically but not isomorphically embedded in $A_2$.} and the causal wedge of $A$ is $a_1 \cup a_2$.

\begin{figure}
    \centering
    \includegraphics[width=140mm]{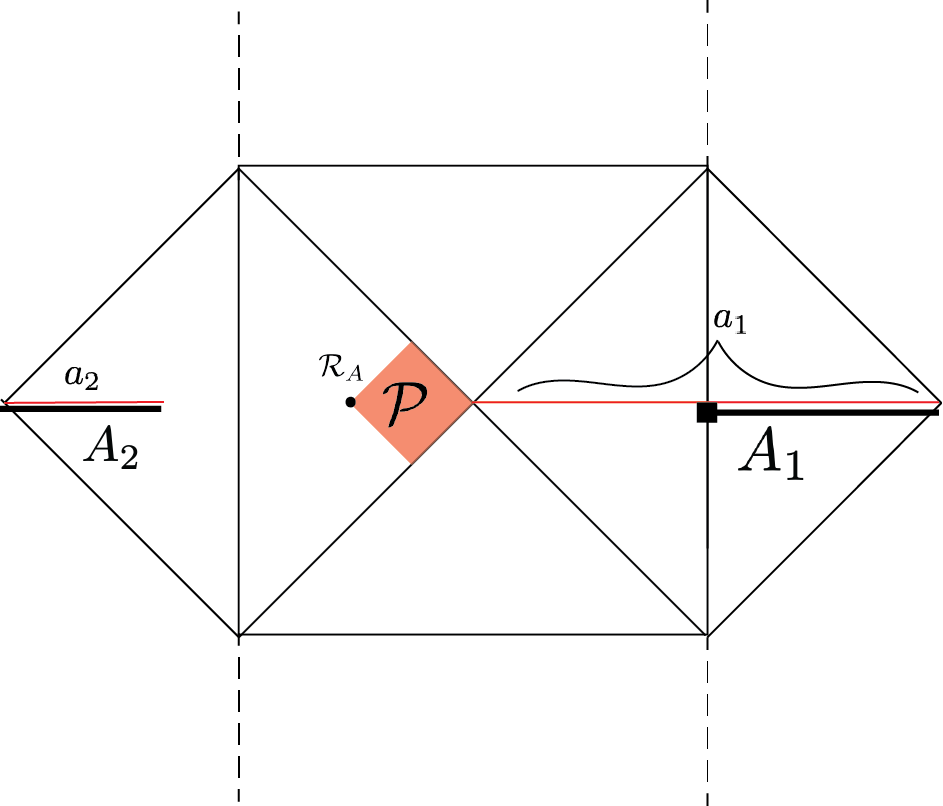}
    \caption{We consider the region in the dual theory $A = A_1 \cup A_2$. In this figure, $A_1$ and $A_2$ are overlaid on the dual picture. The relevant regions in the gravity theory $a_1$ and $a_2$ are also shown.}
    \label{fig:regions}
\end{figure}
This set-up was considered in \cite{AMM} and they found that to leading order in $c\beta/\phi_r$, the region $a_{1\cup 2}$ consisted of 
\begin{align}
a_{12} \equiv &\left \lbrace (w^+, w^-): w^+ \geq \frac{\beta}{4\pi \phi_r}S'(a_1 \cup a_2), \ w^- \leq -\frac{\beta}{4\pi \phi_r}S' (a_1 \cup a_2) \right \rbrace \cup a_2 \nonumber \\ 
&S'(a_1\cup a_2) \equiv  \frac{\partial }{\partial W_1^-}S(a_1\cup a_2) = -\frac{\partial }{\partial W_1^+}S(a_1 \cup a_2) < 0, \label{eqn:dI}
\end{align}
where $S'(a_1\cup a_2)$ is the null derivative of the bulk entropy under an increase in size of $a_1 \cup a_2$.
Note that, since $W_{1}^{+} = W_{1}^{-}$, the derivative should be carefully defined: we take the derivative with the null coordinate of the end-point of $a_{2}$ \emph{before} imposing this equality.
One derives the sign of $S'$ by noting that $S'(a_1\cup a_2) = - I' (a_1:a_2)$ where $I(a_1:a_2)$ is the mutual information between $a_1$ and $a_2$. The derivative of the mutual information has a sign that is fixed by strong sub-additivity.

\subsection{Constructing the Bulk States in JT Gravity}
We now construct states of the bulk field theory coupled to JT gravity where the field theory has been acted on by its own cocycle unitary. We show that to leading order in $c\beta/\phi_r$, the negative energy in this state is enough to bring all excitations in the peninsula into view.

As discussed in the Introduction, we imagine a low energy observer that will act with the cocycle on the causal wedge algebra.  In this case, we want to act with the cocycle unitary introduced in Sec. \ref{sec:cc}, $u_s(\Omega, a_1 \cup a_2) = \sigma_{a_1}^{is} \sigma_{a_2}^{is}\sigma_{a_1\cup a_2}^{-is}$. Remember that in this notation $a_{12} \supsetneq a_1 \cup a_2$. We thus consider states of the bulk field theory of the form
\begin{align}\label{eqn:psis}
\ket{\Omega_s} = u_s(\Omega, a_1 \cup a_2)\ket{\Omega}.
\end{align}

As we discussed above, these states have multiple shocks, one at each entangling surface. Since $a_1$ is the whole left Rindler wedge, and $a_2$ corresponds to an interval in the right Rindler wedge, the state $\ket{\Omega_s}$ has four relevant shocks, two associated to region $a_2$ and two associated to $a_1$, with one left-moving and one right-moving shock for each interval. Note that technically there are also shocks off at $\mathscr{I}^{\pm}$ but their magnitude is zero by calculation.

\begin{figure}
    \centering
    \includegraphics[width=140mm]{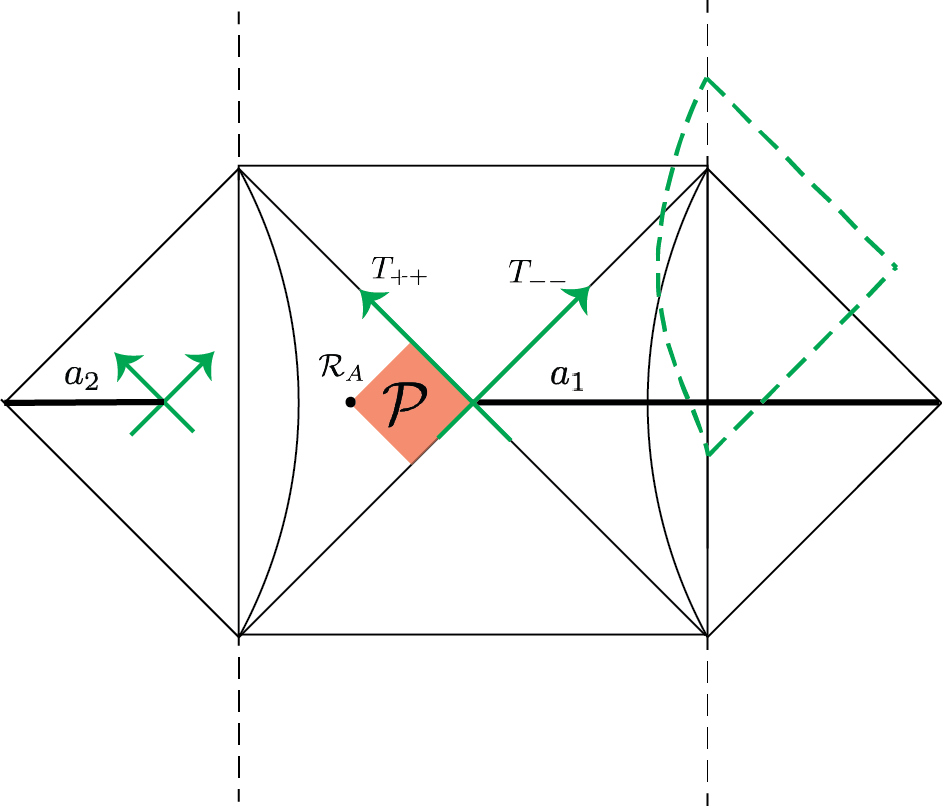}
    \caption{We act with the two-sided unitary $u_s(\Omega, a_1 \cup a_2)$ on the vacuum as in \eqref{eqn:psis}. This has the effect of moving the right boundary particle's trajectory to the dashed green line so that the whole peninsula $\mathcal{P}$ is in the past of the right boundary particle's worldline, to leading order in large $c$. The shocks are labeled by the green arrows. For the purposes of seeing the peninsula, the left-most shocks are irrelevant, although they will have an effect on the left boundary particle's trajectory, which is not shown. Also note that the regions denoted $a_1, a_2$ are the $s=0$ positions of the regions. Note that even though $A_1$, $A_2$ are defined as fixed regions in the baths, their coordinate positions change in the picture after we act with $u_s$. We caution that the main text only works to leading order perturbatively and so we have exaggerated these effects in this figure.}
    \label{fig:jt-eg}
\end{figure}

We are interested in the boundary particle positions in the geometry with these shocks inserted. As described in the previous section, at the bifurcation surface, there are two shocks, one shock of size $\braket{T_{++}} =- (e^{-2 \pi s}-1)S'(a_1 \cup a_2) $ and another of size $\braket{T_{--}} = -(e^{2\pi s}-1) S'(a_1 \cup a_2)$, with $S' < 0$ as defined in \eqref{eqn:dI}. For either sign of $s$, one shock is negative (and bounded in size as $|s| \to \infty$) while the other is positive and growing exponentially. At large enough $s$, the positive energy shock will affect the ability of the negative energy shock to transmit information from the peninsula to the boundary. We will find, however, that we do not need to go to such large $s$. In fact, as we will see, the maximum $s$ we will need is of order $\log(c)$, which does not scale with $\phi_r/\beta \sim 1/G_N$ in any way.

To compute the boundary particle positions, we use the $SL(2,\mathbb{R})$ charge formalism introduced in \cite{MSY-1}. Jackiw-Teitelboim gravity coupled to matter can be written as three systems corresponding to the left and right boundary particles and the matter system. These three systems are only coupled through an $SL(2,\mathbb{R})$ gauge symmetry, which is present because only the relative position of particles in the fixed $AdS_2$ background is physical. The symmetry generators of $AdS_2$ can be simply expressed as vectors in the embedding space formalism for Lorentzian $AdS_2$ \cite{MSY-1}, where $AdS_2$ takes the form of a hyperboloid in $\mathbb{R}^{2,1}$ given by the equation
\begin{align}
-Y_{-1}^2 - Y_0^2 +Y_1^2 = -1.
\end{align} 
 
Each system has its own gauge charge $Q^a_{l,r,m}$ for the left and right boundary particles and the matter systems respectively, corresponding to each of the symmetries of $AdS_2$ with $a$ running over the embedding space coordinates $a = -1,0,1$. The gauge constraints tell us that these charges must add to zero $Q^a_l + Q^a_r + Q^a_m=0$. 

As discussed in \cite{MSY-1, LMZ}, in the large $\phi_r$ limit, the boundary particle trajectories can be expressed in terms of an embedding space vector $Y_{l,r}^a$ that obeys the equation $\lim_{\epsilon \to 0} Q_{l,r} \cdot (\epsilon Y_{l,r}) = \mp \phi_r$, where the different signs correspond to left and right boundary respectively. This describes a hyperbolic trajectory with end-points that are null-separated from the point $Q^a/\sqrt{-Q^2}$ in $AdS_2$. Thus, we can think of these charges as describing the position of the causal  ``surface'' (point in $1+1$-d). To assess whether the shocks have successfully revealed the peninsula, we need to compute the new gauge charges upon inclusion of the shocks. Note that of course our gauge charges are not gauge-invariant, but we will really be interested in the relative position between the future end-point of the boundary particle and the edge of the peninsula, which is gauge-invariant.

When we introduce matter excitations, we need to decide how to dress them. This amounts to deciding which charge $Q_{l,r}$ to subtract $Q_m$ from. In the states we are considering, we choose to dress the shocks associated to $a_1$ to the left boundary and the shocks in $a_2$ to the right boundary. In terms of the matter stress tensor these charges are
\begin{align}
Q^a_m = 2\pi \int_{\Sigma} n^{\mu} T_{\mu \nu} (\zeta^a)^{\mu}
\end{align}
where the $\zeta^a$ are the three Killing vectors of $AdS_2$. These are listed in appendix \ref{app:jt}. 

Since we only care about the right boundary particle's trajectory in determining whether matter can traverse from the peninsula to the right black hole, we only need to compute the contribution to the matter charge from the shocks at the black hole bifurcation surface.  For the finite temperature eternal black hole, the charges obey $Q_l^a = -Q_r^a = (\sqrt{\mu},0,0)$ and thus the bifurcation surface sits at $Y^a = (1,0,0)$, which is where the two right matter shocks will intersect. Here $\sqrt{\mu} = \frac{2\pi \phi_r}{\beta}$ is related to the mass of the black hole.

The matter charge for this configuration is easily computed to be 
\begin{align}
Q^{-1}_m = 0,\ \ Q^0_m(s) = \frac{1}{2}(p_-(s)-p_+(s)),\ \ Q^1_m(s) = \frac{1}{2}(p_-(s) + p_+(s))
\end{align}
where we used equation \eqref{eqn:energyshocks} of the previous section to write
\begin{align}
p_{\pm}(s) =\pm (e^{\mp 2\pi s}-1)\partial_{\pm} S(a_1\cup a_2) = (1 - e^{\mp 2\pi s}) S'(a_1\cup a_2),
\end{align}
where we again remind the reader that $S' < 0$.
 
In $w^{\pm}$ coordinates, these shocks move the causal surface to the point $w_{cs}^{\pm}$ determined by the equations
\begin{align}
Y^{-1} &= \frac{1-w_{cs}^+ w_{cs}^-}{w_{cs}^+ w_{cs}^- +1} = \frac{\sqrt{\mu}+Q_m^{-1}}{\sqrt{-Q^2}} = \frac{\sqrt{\mu}}{\sqrt{\mu-p_-(s) p_+(s)}} \nonumber \\
Y^\pm \equiv \frac{ Y^0 \pm Y^1}{2} &= \frac{w_{cs}^\pm}{1 + w_{cs}^+ w_{cs}^-} = \frac{Q_m^0 \pm Q_m^1}{2\sqrt{-Q^2}} = \pm \frac{p_{\mp} (s)}{2\sqrt{\mu - p_+(s) p_- (s)}}.
\end{align}
Solving for $w_{cs}^{\pm}$, we find
\begin{align}\label{eqn:causalsurface}
w^{\pm}_{cs} = \pm \frac{\sqrt{\mu} - \sqrt{\mu -p_-(s) p_+(s)}}{p_{\pm}(s)},
\end{align}
which, for $p_-(s) p_+(s) \ll \mu \sim \phi_r^{-2}$ gives
\begin{align}
w^{\pm}_{cs} \approx \mp \frac{1}{2} p_{\mp}(s)/\sqrt{\mu} = \pm (1 - e^{\pm 2\pi s}) \frac{\beta}{4\pi \phi_r} S'.
\end{align}

For positive $s$, we see that the causal surface gets shifted in the positive $w^-$ direction, thus allowing one to see more of the peninsula. In the (very) large $s$ limit, however, we see that $w^-_{cs} \to 0$. In this limit, the positive (and exponentially growing in $s$) energy of the left-moving shock takes over and closes off our ability to see the peninsula. Thankfully, our goal is more modest in that we are just trying to see to within a Planck distance of the edge of the peninsula. To achieve this, we see that $s$ needs to be of order $\log c$. 

At this value of $s$, $w_{cs}^+$ takes the form
\begin{align}
w_{cs}^-(s_* \sim \log c) = - \frac{\beta}{4\pi \phi_r} S'(a_1\cup a_2) + \mathcal{O}(c^0\beta/\phi_r)
\end{align}
where we used the equation $\sqrt{\mu} = 4\pi \phi_r/\beta$ derived in \cite{MSY-1}, relating $\phi_r/\beta$ to the boost charge of the black hole. Comparing with equation \eqref{eqn:dI}, we see that in these states we can see to within a distance of order the Planck distance from the quantum extremal surface. We should not expect to be able to achieve better resolution than this while staying within the semi-classical approximation. We thus have achieved our goal for this example. 
Finally, notice that \eqref{eqn:causalsurface} also implies that the CC flow with $s<0$ shifts the causal surface to the past, allowing boundary observers to affect the peninsula.
One can generalize this set-up to higher dimensional AdS/CFT and we will do so in the following section.

%%%%%%%%%%%%%%%%%%%%%%%%%%%%%%%%%%%%%%%%%%%%%%%%%%%%%%%%%%%%%%

\section{Seeing the Entanglement Wedge in the General Perturbative Case} \label{sec:gen}

In this Section, we attempt to argue in a more general setting that the backreaction from a unitary confined to the causal wedge in some original bulk state can reveal the peninsula.

The setup here is an asymptotically $AdS_{d+1}$ spacetime gravitationally coupled to a local bulk QFT. The dual theory is a holographic CFT living on a $d-$dimensional boundary sphere.
 Let us call the boundary region under consideration $A$. We will choose $A = \cup_{r=1}^n A_r$ to be a union of $n$ boundary spherical caps.\footnote{We expect the results to generalise to cases in which one or more of the $A_r$s is an entire CFT.}
 We set the AdS radius to $\ell = 1$.
We use the notation $\ket{\psi}$ to interchangeably mean a state of the bulk quantum fields and the boundary state.
 
 With this choice of $A$ there is a natural class of boundary states whose causal and entanglement wedges agree --- they are a union of $n$ Rindler wedges.
 These states are known as split vacua; they are purifications of the reduced state $\bigotimes_{r=1}^n \sigma_r$ on $A$, where $\sigma_r$ is the vacuum restricted to $A_r$.
 We pick a particular purification and call it $\ket{S_\Omega}$.
 The wedges agree because the reduced state is $n$ uncorrelated copies of the vacuum, so that the entanglement wedge is the union of the vacuum entanglement wedges of all the $A_r$s; and the vacuum entanglement wedge of a spherical cap is the same as its causal wedge.
 For $n=1$, an example of a split vacuum is just the global vacuum.
 
 We take the state of interest to be an excited state $\ket{\psi}$ that is `close' to the vacuum within its outermost wedge so that it  has a non-trivial Planck-scale peninsula.
More precisely, for some parameter $c$ such that $1 \gg c G_N \gg G_N$, we require that within the outermost wedge the bulk stress-energy and entanglement entropy shape derivatives in $\ket{\psi}$ scale with $c$ and the bulk metric is $g_\psi = g_{AdS} + h$, $h = \mO(cG_N)$.
 As a result, the proper size of the peninsula scales as $c G_{N}$ where we take the limit $1 \gg c G_N \gg G_N$.
The causal wedge of $A$ in the states we consider can always be written as a union of wedges $\mathcal{W}_C [A_r]$ with Cauchy slices $a^\mathcal{C}_r$, whose boundary $\partial a^{\mathcal{C}}_r$ is homologous to $A_r$ ($a^{\mathcal{C}} = \cup_{r=1}^{n} a^{\mathcal{C}}_r$); this is equally true of the outermost wedge, by definition.
Thus, the peninsula is also a union of connected regions each of which is associated to some boundary region $A_r$.
 
\begin{figure}%
    \centering
{\includegraphics[width=.365\columnwidth]{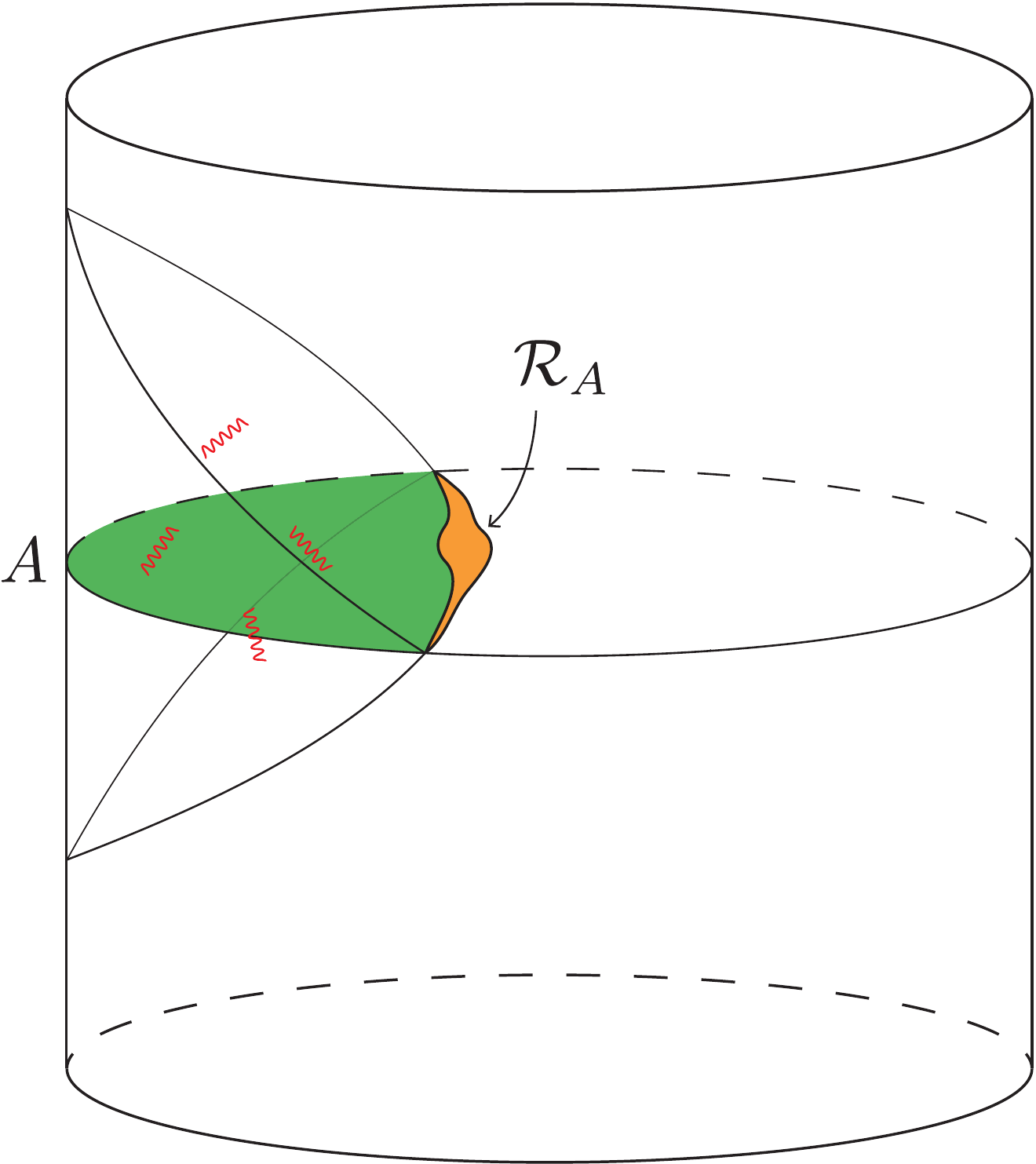} }%
    \qquad
{\includegraphics[width=.4\columnwidth]{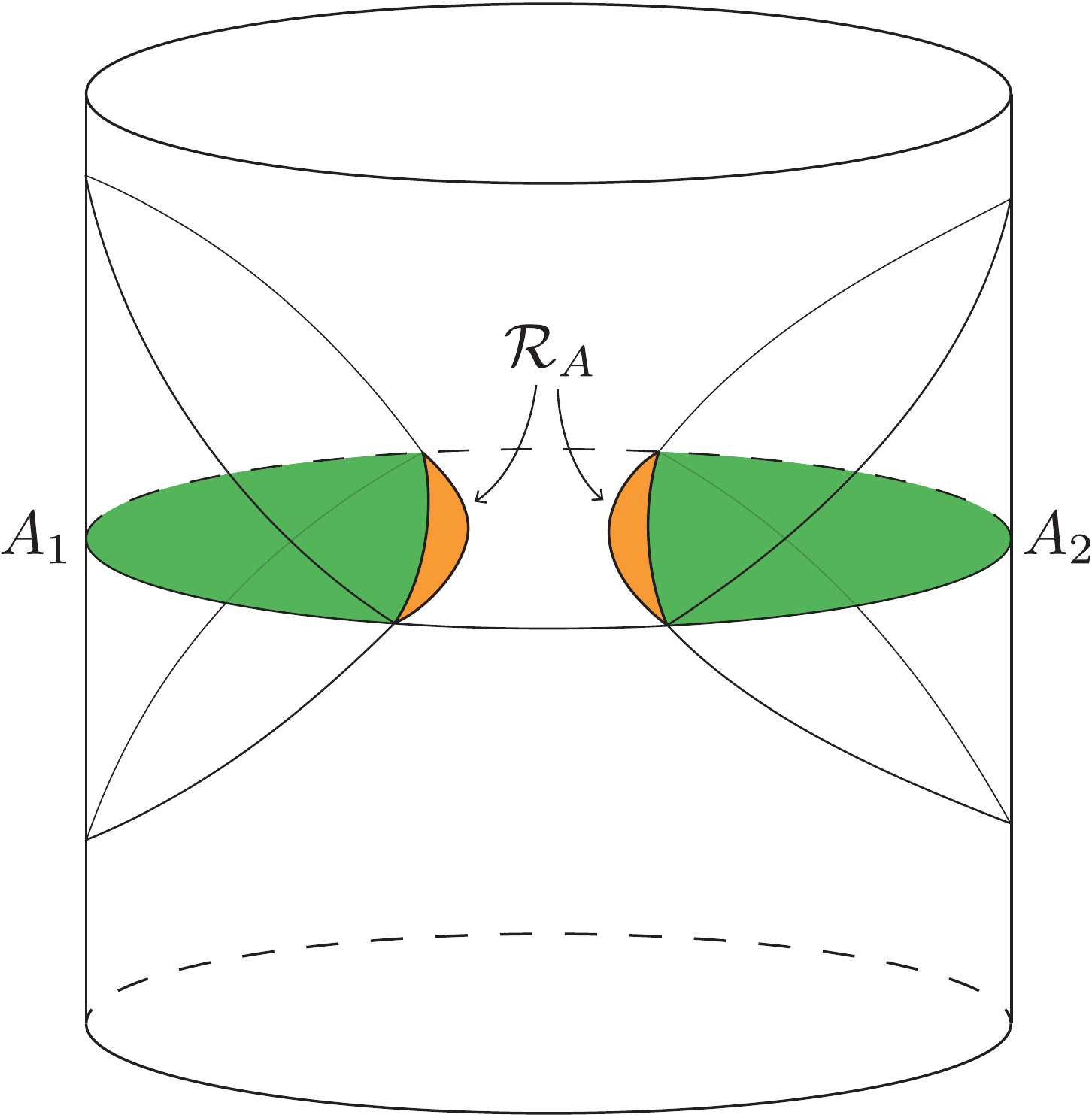} }%
    \caption{Left: When the boundary region is a spherical cap, the entanglement wedge and the causal wedge agree in the vacuum, but small perturbations to the state with $\langle T_{\mu\nu}\rangle \sim c$ will generate a small gap between them. The green region is the a slice of the causal wedge and the orange region is bounded by the causal surface and the RT surface $\mathcal{R}_{A}$ and its domain of dependence is $\mathcal{P}[A]$.  Right: In pure $AdS$, a boundary region $A = A_{1}\cup A_{2}$, a union of two disjoint spherical caps is shown. Due to the bulk entanglement between the disjoint components of the causal wedge, the outermost quantum extremal surface $\mathcal{R}_{A}$ bulges inwards from the causal surface by an amount proportional to $cG_{N}$. Note that if $A$ is large, the RT surface will be different from $\mathcal{R}_{A}$.}%
    \label{fig:examples}%
\end{figure}

 There are intuitively two reasons a non-empty peninsula can exist in this setup: entanglement and energy.
 We mention primitive examples where each effect is dominant; the general case has a combination of the two.
 The first example is a higher-dimensional version of the setup in Section \ref{sec:peninsula}, where we take the number of components $n > 1$, the bulk theory to have $\mO(c)$ light fields and the `excited' state $\ket{\psi}$ to be but the global vacuum $\ket{\Omega}$.
 In this case, while the union of the causal surfaces of each $A_r$ continues to be a \emph{classical} extremal surface, it fails to be a \emph{quantum} extremal surface, because of the bulk entanglement between the various Rindler wedges.
 Since the theory has $\mO(c)$ light fields, the entanglement derivatives scale with $c$,\footnote{We can take this to be the definition of the bulk theory having an $\mO(c)$ light fields in the case when the bulk theory is strongly interacting.} and the peninsula has size $\mO(cG_N)$ (Fig. \ref{fig:examples}).
 
 One can also have a peninsula of size $cG_N$, for $n = 1$ in any bulk theory, if the state $\ket{\psi}$ is obtained by the action of a local unitary well within the causal wedge that creates bulk stress energy $\langle T_{\mu\nu} \rangle_\psi = \mO(c)$. 
 The bulk metric of this state then differs from pure AdS by $h = \mO (c G_N)$, and again this will cause there to be an $\mO(cG_N)$ gap between the causal and outermost extremal surfaces.
 Clearly, in general, both these effects can contribute.

In this setup, our goal is to define some one-parameter family of unitaries, which we will call $\mathfrak{u}_s(\psi,a^{\mathcal{C}})$  that have the following two properties: 
\begin{itemize}
    \item \textbf{Property 1}: The stress energy tensor of the quantum fields in the state $\mathfrak{u}_s(\psi,a^{\mathcal{C}})\ket{\psi}$ agrees with the stress energy of the cocycle on a pure AdS background to leading order in the $cG_N$ expansion.
    \item \textbf{Property 2}: The quantum state and geometry of $\mathcal{P}[A]$ as well as the region space-like to the outermost wedge are preserved by the action of the unitary $\mathfrak{u}_s(\psi,a^{\mathcal{C}})$. 
\end{itemize}
Defining an operator that satisfies these two properties is somewhat subtle for two reasons.
The first property is subtle because we understand the Connes cocycle flow operator only in the vacuum $AdS$ background, where there is a boost symmetry; but we want an operator $\mathfrak{u}_s (\psi, a^\mathcal{C})$ that has a similar effect on the \emph{perturbed} geometry.
We will deal with this by explicit construction; this will not be a unique construction, but we only need existence not uniqueness.
The second property is subtle because of diffeomorphism-invariance in the bulk, and the fact that the Connes cocycle flow is generically localised to a full Cauchy slice within the causal wedge.
Operators deep in the bulk carry a gravitational dressing so as to not break diffeomorphism-invariance; and we essentially need to make sure that the dressing also commutes with the signal.
We will take care of this by being explicit about the dressing.
We will turn to these questions again in Sec. \ref{ssec:construction}, where we attempt to propose such an operator.

When $n>1$ and when $\psi = \Omega$ is the vacuum, an operator that satisfies both of these properties is merely the bulk CC flow operator $u_{s}(\Omega, a)$ of Sec. \ref{sec:cc}.\footnote{In pure $AdS$, $a = a^{\mathcal{C}}$.} To find the corresponding semiclassical state one needs to solve for the metric perturbation $h_{\mu\nu}$ that the stress tensor distribution in Eq. \eqref{eqn:shocks} causes. Given the simplicity of this setup, we assume there exists solutions to $h_{\mu\nu}$ such that the change in the geometry is fully contained away from $\mathcal{P}[A]$.\footnote{One explicit way to do so is to construct bulk-to-bulk Green's functions associated with the inhomogenous linearized Einstein field equations whose support vanishes outside of $\mathcal{W}_{A}[C]$, but we will leave this to future work.} This would amount to a particular gravitational dressing of the QFT operator $u_{s}(\Omega, a)$.\footnote{In general, we expect to be able to perturbatively turn on backreaction on the same stress tensor distribution in different ways. In \cite{Bousso:2020aa}, certain bulk states where discussed where the same stress tensor shocks sit at the quantum extremal surface. Even though the stress tensor distribution of these states agree in the $G\to 0$ limit with that of Sec. \ref{sec:cc}, these states do change the state beyond the causal wedge of the original geometry and therefore do not respect our property 2.}

In cases where the initial metric differs from pure AdS, explicitly constructing such a unitary is difficult as we will discuss in sub-section \ref{ssec:construction}. Before discussing those subtleties, however, we will demonstrate why the two itemized properties above are sufficient for seeing $\mathcal{P}[A]$.

\subsection{\textbf{Properties 1 \& 2} are sufficient} \label{ssec:calculation}

Our argument will be to compute the null shift experienced by a bulk geodesic leaving the future tip of the domain of dependence $D(A)$ of some boundary region $A$ and traveling back in time in the flowed states, $\ket{\psi_s} = \mathfrak{u}_s(\psi,a^{\mathcal{C}}) \ket{\psi}$, assuming that properties 1 \& 2 described above both hold. We will find that, to leading order in $cG_N$, the null shift will be of precisely the right value so that geodesics leaving from the so-called ``peninsula'' and traveling toward the future can make it into $D(A)$.

\subsubsection*{Solving for the Causal Wedge in $\ket{\psi_s}$}
To understand the causal wedge in $\psi_s$, we can examine one of the component entanglement wedges, $a_i$. For simplicity, one can work in Poincare coordinates, with the region $A_i = \{ x^+ > 0, x^- <0 \}$, so that the component of $A$'s RT surface associated to $a_i$ lies perturbatively close to $(x^+ = 0, x^- =0,y^i,z)$. The metric in $\psi_s$ is then
\begin{equation}
    ds^2 = \frac{-dx^+ dx^- + d \vec{y}^2 + dz^2}{z^2} + h^s_{\mu\nu} dx^\mu dx^\nu,
    \label{eqn:metric}
\end{equation}
with $h^s_{\mu\nu} \sim c G_N$.

The null shift experienced by a geodesic in the $\psi_s$ spacetime can be found perturbatively by solving for the position of the causal surface. Similarly to solving for the RT surface position, the causal surface can be found by solving for a co-dimension two surface whose null expansion reaches zero at infinity as one evolves the surface in a null direction toward the boundary. One can set up null vectors which generate the congruence leaving the causal surface. We call these generators $k$ and $\ell$, such that $k\cdot \ell = 1$ and both vectors point toward the boundary region $A$. By integrating Raychaudhuri's equation $\dot{\theta^{(k)}} = \frac{-1}{d-2} (\theta^{(k)})^2-\sigma_k^2 - 8\pi G T_{\mu\nu}k^{\mu}k^{\nu}$ and setting $\lim_{\lambda \to \infty}\theta^{(k)}(\lambda) = 0$, we find
\begin{align}\label{eqn:causalsurface-gen}
    \theta^{(k)}_{CS} = \int_0^{\infty}d\lambda\, \braket{T_{\mu\nu}}k^{\mu}k^{\nu} + \mathcal{O}(G_N^2)
\end{align}
where we have used the fact that the shear-squared and expansion-squared terms are higher order in the $cG_N$ expansion. An exactly analogous equation holds for the $\ell$-direction.

To solve for the null shift of the causal surface, we can expand perturbatively about the background AdS causal surface, which in Poincare coordinates sits at $x^+ = x^- = 0$. We can describe the causal surface in terms of embedding functions $x^{\mu} = \delta X_c^{\mu}(y^i,z) = \delta X_c^- \ell^{\mu} - 2z^2 \delta X_c^+ k^{\mu}$ where we pick $k_{\mu} = \delta^-_{\mu}$ with $\ell$ such that $k\cdot \ell = 1$.

In terms of these embedding functions for the causal surface, equation \eqref{eqn:causalsurface-gen} can be written as
\begin{align}\label{eqn:CSexpansioneq}
    \frac{k_{\mu}}{\sqrt{H}} \partial_{\alpha}\left(\sqrt{H} H^{\alpha \beta} \partial_{\beta} X^{\mu}(y^i,z)\right) + \Gamma^{\mu}_{\rho \sigma}H^{\alpha \beta}\partial_{\alpha}X^{\rho}\partial_{\beta}X^{\sigma} + 8\pi G z^2 \int_0^{\infty} dx^+ \, T_{++}(x^-=0,x^+,y^i,z)=0
\end{align}
where $H_{\alpha \beta}$ is the induced metric of the surface (and $H$ the determinant) and we traded the affine parameter $\lambda$ defined by $k^{\mu} = \left(\frac{d}{d\lambda}\right)^{\mu}$ for the $x^+$ coordinate. Furthermore, in the integral of the stress tensor, we used the fact that $T_{++}(\lambda) \approx T_{++}(x^-=0,x^+(\lambda))$ to leading order in the $cG_N$-expansion.

Perturbatively expanding \eqref{eqn:CSexpansioneq} in both $X^{\mu}$ and the metric, we get that to leading order the null shift obeys the differential equation 
\begin{align}\label{eqn:shifts-0}
\mathcal{D}\, \delta X^-_c (s) &=  8\pi G \int_0^{\infty} dx^+ \,\braket{T_{++}(x^-=0,x^+,y^i,z)}_{\psi_s} + 4z^2\partial_- h^s_{++}(x^-=0,x^+=0,y^i,z)
\end{align}
where $\mathcal{D}$ is the differential operator $\mathcal{D} = \partial_z^2 + \sum_{i=1}^{d-2} \partial_{y^i}^2 + \frac{1-d}{z} \partial_z$.

By \textbf{Property 1}, since the stress of the quantum fields in the $\psi_s$ state are assumed to be that of the flowed state on pure AdS to leading order, then the $x^+$-integral will pick up the shock at $x^+ = x^- = 0$. Furthermore, using the formulae for the integrated null energy in the flowed states in Section \ref{sub-summary}, we find that the null shift of the causal surface to leading order in $cG_N$ is given by
\begin{align}\label{eqn:shifts}
 \mathcal{D} \, \delta X^-_c (s) &=  (e^{-2\pi s}-1) 4G_N z^{d-1} \frac{\delta S_{bulk}}{\delta X^+(y,z)} \nonumber\\&\qquad + e^{-2\pi s}8\pi G_N \int_0^{\infty} dx^+ \,\braket{T_{++}(x^-=0,x^+,y)}_{\psi_{s=0}} + 4z^2\partial_- \left(h_{++}\right)_s
\end{align}
At $s_* \sim \log c$, the integral over the stress tensor in the second line of \eqref{eqn:shifts} is order $G_N$ and not order $cG_N$ and so we can ignore it. We thus get 
\begin{align}\label{eqn:deflect}
&\mathcal{D} \delta X^-_c (s_*) \approx  -4G_N z^{d-1}\frac{\delta S_{bulk}}{\delta X^+(y,z)} + 4z^2\partial_- h^s_{++}  +\mathcal{O}(c^0 G_N).
\end{align}

By \textbf{Property 2}, the metric in the peninsula is unaffected by the unitary flow. Since equation \eqref{eqn:deflect} comes from examining the expansion of the causal surface, the term $\partial_- h_{++}^s$ should really be evaluated on the causal surface in the $\psi_s$ spacetime. By continuity of the metric in the $\ket{\psi_s}$ state, we conclude that the $\partial_- h_{++}^s$ term in \eqref{eqn:deflect} is actually independent of $s$. This will be important momentarily.

\subsubsection*{Solving for the Outermost Wedge in $\ket{\psi_s}$}
We now show that the solution $\delta X^-_c(s_*)$ to the differential equation in \eqref{eqn:deflect} is precisely the position of the RT surface in the original, un-flowed geometry. This means that in our states at large $s_* \sim \log c$ the geodesic made it from the future tip of $D(A)$ to the quantum extremal surface. This was our goal.

To see that the solution to \eqref{eqn:deflect} corresponds to the null shift relative to the (split) vacuum of the RT surface in the un-flowed geometry, we need to solve for the quantum extremal surface equation. The position of the quantum extremal surface can also be described via embedding functions $\delta X_{RT}^{\mu}(y,z) = \delta X^-_{RT} \ell^{\mu} -2z^2 \delta X^+_{RT} k^{\mu}$, where again the factor of $-2z^2$ comes from the normalization of $k^{\mu}$ so that $k \cdot \ell = 1$. The quantum extremal surface equation says that 
\begin{align}
 \frac{\theta^{(+)}(y,z)}{4G_N} + \frac{k^{\mu}}{\sqrt{H}}\frac{\delta S_{bulk}}{\delta X^{\mu}(y,z)} = \frac{\theta^{(-)}(y,z)}{4G_N} + \frac{\ell^{\mu}}{\sqrt{H}}\frac{\delta S_{bulk}}{\delta X^{\mu}(y,z)} = 0
\end{align}
where $\theta^{(\pm)}$ is the classical expansion in the $\pm$-direction of a candidate RT surface. 
Following the same steps as for the causal surface, we expand perturbatively in the embedding functions $X^{\mu} \to X^{\mu} + \delta X^{\mu}$ as well as the metric, and find
\begin{align}\label{eqn:RT}
\mathcal{D}\, \delta X^-_{RT}(z,y) = -4G_N z^{d-1}\frac{\delta S_{bulk}}{\delta X^+(z,y)} + 4z^2\partial_- h_{++}.
\end{align}

\subsubsection*{The Result}
Subtracting equations \eqref{eqn:RT} and \eqref{eqn:shifts}, we find that
\begin{align}
    \delta X_{RT}^- - \delta X_c^- (s) &= - e^{-2\pi s}\ 8\pi G_N\ \mathcal{D}^{-1} \left( \int_0^\infty \langle T_{++} \rangle_{s=0} + \frac{z^{d-1}}{2\pi} \frac{\delta S_{bulk}}{\delta X^+} \right) \nonumber\\
    &= e^{-2\pi s}\left( \delta X_{RT}^- - \delta X_c^- (s=0)\right) \xrightarrow{s = s_* \sim \log c} \mathcal{O} (c^0 G_N)
\end{align}
This is precisely what we wanted to show.

A similar calculation of $\delta X_{RT}^+-\delta X_c^+$ shows that this component of the gap is proportional to $e^{2\pi s}$.
This means that for $s < 0$ $D(A)$ can bring $\mathcal{P}[A]$ into the future of $D(A)$ instead of the past, i.e. we can affect the peninsula instead of seeing it.

\subsubsection*{Disaggregating the Effects}

At given value of $s$ used in the cocycle flow, the null shift experienced by a geodesic leaving from the future tip of $D(A_r)$ is given in equation \eqref{eqn:shifts}. On the right hand side of this equation there are two contributions. The first term is from the shock of energy at the causal surface, and the second is due to the pre-existing energy in the causal wedge of $A_i$ falling across the horizon. The $s$-dependence of this last term is from a simple boost of the stress energy.

\begin{figure}[h]
    \centering
    \includegraphics[width=140mm]{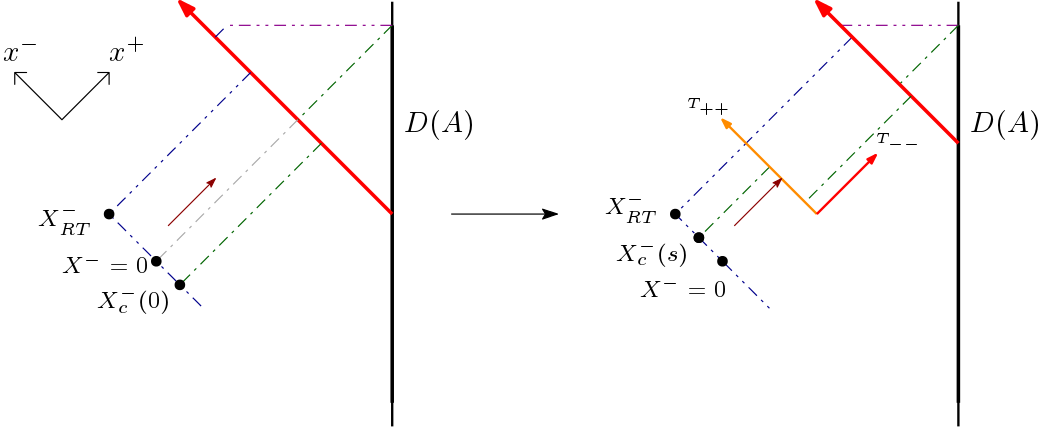}
    \caption{Schematic representation of a simple case where the gap is sourced by both energy --- the red shock --- and entanglement --- not pictured. In the new state, the energy is boosted up, so that it causes less of a time-delay; and there is an infalling negative energy shock, creating a time-advance to close the entanglement-sourced part of the gap.
    This is not a Penrose diagram; time-delays and advances are shown as simple changes in position.
    This is only the leading-order behaviour.
    Green lines mark causal horizons, red lines mark positive energy shocks, orange lines mark negative energy shocks, and the brown line is a signal leaving the peninsula.}
    \label{fig:pert-geom}
\end{figure}

A useful way to think about these two effects and the distinction between them is to remember that the gap between the causal and entanglement wedges is created by two effects, entanglement and energy.
Energy sources the gap in an obvious way in that the energy of the bulk excitations backreacts on the metric and bends null lines.
The separation between the causal and entanglement wedge is also sourced by bulk entanglement in that the location of the quantum extremal surface is determined by bulk entropy derivatives --- this was the source of the entire gap in Section \ref{sec:peninsula}.
Thus, we may heuristically think of the negative energy shock and the boosting of the bulk energy as closing these two different sources of the gap.
This is illustrated in figure \ref{fig:pert-geom}.

Another perspective, that furnishes intuition for the importance of CC flow, is as follows.
To leading order, the null separation between the HRT and causal surfaces in the $X^-$ direction is proportional, by Einstein equations, to the quantum half-averaged null energy (QHANE) of the future horizon $\sim \int_0^\infty T_{++} dx^+ + \frac{\delta}{\delta X^+} S$.
The effect of CC flow is to exponentially damp this quantity, sending it to $0$ and saturating the so-called QHANEC at large $s$.
Another fascinating point is that $\mathcal{W}_C[A] \subseteq \mathcal{W}_E[A]$ was shown in \cite{Akers:2016aa} to imply the \emph{boundary} QHANEC; here, assuming \textbf{Properties 1 \& 2} we are saturating $\mathcal{W}_C[A] \subseteq \mathcal{W}_E[A]$ by saturating \emph{bulk} QHANE; further in section \ref{sec:bd} we show that the effect is the same as the \emph{boundary} CC flow, which saturates the \emph{boundary} QHANEC.\footnote{Note that, as in Section \ref{sec:peninsula} we are only saturating $\mathcal{W}_C[A] \subseteq \mathcal{W}_E[A]$ in one null direction --- the $x^-$ direction in the calculation above. In the other null direction, at large $s$, $\mathcal{W}_C[A] \subseteq \mathcal{W}_E[A]$ is actually being taken away from saturation, but this is irrelevant for seeing the peninsula.}

\subsection{Does a $\mathfrak{u}_s$ exist that satisfies \textbf{Properties 1 \& 2}?} \label{ssec:construction}
We now turn to the question of whether there is a unitary $\mathfrak{u}_s(\psi,a)$ which satisfies \textbf{Properties 1 \&2}. In Section \ref{sec:peninsula}, we were able to define states in the bulk which neatly meet two important requirements for our states: 1. they solved the JT equations of motion coupled to matter and 2. arose from the action of a well-defined bulk QFT unitary --- the cocycle. Since the state of quantum fields in the so-called peninsula were manifestly untouched, we were able to claim to see the contents of the peninsula ``causally.'' This construction used the fact that quantum fields coupled to JT gravity propagate on a fixed AdS$_{2}$ background. 

In higher dimensions (or for different dilaton potentials in 1+1-d), the situation is a bit trickier. There are two main complications that arise. The first is that it is hard to consistently solve analytically for solutions to the equations of motion with a given stress energy profile. The second issue, which is more serious, is that in general the background metric is affected by moving from some (split) vacuum state to an excited state $\psi$. It is thus no longer clear how to define the cocycle between these two states purely from the bulk. Nevertheless, we now propose a construction that we suspect will work but is somewhat hard to work with.

\subsubsection*{A Possible Construction}
To remind the reader, we consider our boundary region $A$ to be the union of $n$ disconnected spherical caps. We consider an excited state $\psi$ whose outermost quantum extremal surface homologous to $A$ differs from its causal surface by the small parameter $cG_N$. We take $c$ to be large, however, so that we can localize low-energy excitations within the peninsula. Since $cG_N$ is small, we can decompose the $\psi$-state metric in the outermost wedge as $g_{\psi}= g_{AdS} + h$. In the background pure $AdS$ metric, the causal wedge is just a union of $n$ disconnected Rindler wedges, each homologous to a component of $A$. In accordance with Sec. \ref{sec:cc}, we denote a Cauchy slice of  this union by $a = \cup_r a^r$.

The operator of interest is related to a well-defined Connes cocycle operator that acts on pure $AdS$; we begin by defining this vacuum $AdS$ operator and then using it to create the non-vacuum operator that satisfies the two properties.
We suppose that $\psi$ was created by insertions of boundary operators dual to the light bulk fields in a path integral; since states of this form are dense in the low-energy subspace of the CFT, this assumption is without loss of generality. We can then define a state on the pure AdS metric \emph{without gravity} by just acting with these light boundary operators and evolving in the bulk using the free field equations of motion. This defines some quantum field theory state on fixed background AdS, which we denote $\psi_{AdS}$. The cocycle of interest is then just $u_s \left(\psi_{AdS},a\right)$ of Sec. \ref{sec:cc}.\footnote{Note that we haven't violated any diffeomorphism constraints: we turn gravity off only to define the state $\ket{\psi_{AdS}}$, and we use this state only to define the operator $u_s(\psi_{AdS},a)$. The bulk effective theory states that we will now go on to construct will always belong to the Hilbert space associated to the correctly backreacted geometry, as mandated by gravity.}

 Note that this construction is completely well-defined.
However, it only acts on the Hilbert space associated to the pure $AdS$ geometry.
For the pure $AdS$ case, i.e. the higher-dimensional version of Section \ref{sec:peninsula}, the geometry dual to the `excited' state $\ket{\Omega}$ is in fact pure AdS, and the CC flow $u_s (\Omega,a)$ is the operator we were looking for.
We merely need to dress it so that it is localised to a region spacelike to the peninsula in the final geometry.

In the general case, however, the bulk metric will generically be perturbed and we need a further, more speculative, step to define an operator that acts on this perturbed background and is a candidate for $\mathfrak{u}_s(\psi,a^{\mathcal{C}})$.
To achieve this, we can use HKLL to map $u_s \left(\psi_{AdS},a\right)$ to the boundary. This requires breaking the cocycle up into its constituent parts. For example, in Gaussian states of free field theory, this cocycle will look like the exponential of a bi-local operator $H$ where \cite{Arias:2018aa, Faulkner:2017vdd}
\begin{align}
    H = \sum_{i=1}^c \int_{A} dx dy\  \phi^i(x) K^i(x,y)\phi^i(y) + ...\ .
\end{align}
The $...$ includes terms with time-derivatives of the fields. Our prescription is then to map this generator to the boundary and exponentiate it. This defines some boundary operator that is a natural candidate for  $\mathfrak{u}_s(\psi,a^{\mathcal{C}})$, the operator that satisfies \textbf{Properties 1 \& 2}.

We can now use HKLL to map this boundary operator back into the bulk in the metric dual to the excited state $\psi$. Since HKLL kernels are expected to be differentiable in the metric \cite{AAL}, we expect that this operator creates the same stress tensor distribution in the bulk as the cocycle operator, up to $\mO (cG_{N})$ corrections.
In the absence of bulk gravity the $\mathfrak{u}_s(\psi,a^{\mathcal{C}})$ unitary on the $\psi$-metric commutes with any operators outside the causal wedge of $A$.

Another way to map $u_s \left(\psi_{AdS},a\right)$ to an operator that can act on a perturbed metric is to use the stress tensor.
To first order, the operator $e^{i \int \sqrt{g_{AdS}} h \cdot T}$ maps states on $g_{AdS}$ to $g_{AdS} + h$.
However, it is not a diffeomorphism-invariant operator in that $\int h \cdot T$ and $\int (h + \mathcal{L}_\xi g_{AdS}) \cdot T$ agree in correlation functions only up to contact terms.
So, we must pick a specific representative $g_{AdS} + \hat{h}$ of the metric dual to $\psi$.
We make the choice in which every operator in the causal wedge $\mathcal{W}_C[A]$ has vanishing Dirac brackets with space-like separated operators; in other words, $\hat{h}$ is the metric in a gauge where every bulk point in $\mathcal{W}_C[A]$ is `dressed' to $D(A)$.
The operator $\mathrm{u}_s (\psi,a^{\mathcal{C}}) = e^{i \int_a \hat{h} \cdot T} u_s \left(\psi_{AdS},a^{\mathcal{C}}\right) e^{-i \int_a \hat{h} \cdot T}$ is then another candidate for the operator of interest; and it is restricted to the causal wedge in the perturbed metric because of our specific choice of gauge.
To see that this causes only an $\mO(cG_N)$ shift in the stress energy distribution of the flowed state, we merely need the fact that the commutator $[T_{\mu\nu} (x'), T_{\alpha \beta} (x)]$ does not scale with $1/G_N$.
This is because
\begin{align}
    \mel{\psi}{\mathrm{u}_s (\psi,a^{\mathcal{C}})^\dagger\ T_{\alpha\beta} (x)\ \mathrm{u}_s (\psi,a^{\mathcal{C}})}{\psi} &= \mel{\psi_{AdS}}{u_s (\psi_{AdS},a)^\dagger\ T_{\alpha\beta}(x)\ u_s (\psi_{AdS},a)}{\psi_{AdS}} \nonumber\\ & \ + i \int_{a} dx' \hat{h}^{\mu\nu} (x') \mel{\psi_{AdS}}{u_s (\psi_{AdS},a)^\dagger [T_{\alpha\beta}(x), T_{\mu\nu} (x')] u_s (\psi_{AdS},a)}{\psi_{AdS}}.
\end{align}
Since $\hat{h} = \mO(cG_N)$, the second term is suppressed unless the expectation value of the commutator diverges as $1/G_N$; we don't expect this to be the case, since the commutator can be written in terms of field theory quantities as e.g. in \cite{Deser:1967zzf}, and therefore $\mathrm{u}_s (\psi,a^{\mathcal{C}})$ creates the right stress tensor distribution at leading order.
Thus, this is also a reasonable candidate for $\mathfrak{u}_s \left(\psi,a^\mathcal{C}\right)$.

To summarise, we have constructed a precise unitary in the pure AdS case, i.e. the higher-dimensional version of Section \ref{sec:peninsula}; and two candidates in the case when the metric is perturbed.
They act in the region spacelike to the peninsula and towards the boundary within bulk QFT; and we assume that it can be ensured that they stay that way once gravitational effects are taken into account by dressing them appropriately.
In the second case, we have not been able to furnish a proof that either candidate doesn't have for example non-integrable divergences somewhere; this is why we feel that this section is more speculative than Section \ref{sec:peninsula}.

\subsubsection*{Towards a Causal Implementation}
There is however an important caveat to be made for the operator constructed here as well as the one used in Section \ref{sec:peninsula}.
Namely, that it may not be possible to HKLL-reconstruct it on the boundary, since the region spacelike to the peninsula in the flowed geometry is not entirely contained in the causal wedge, as can be seen for example in figure \ref{fig:jt-eg}.
Now we discuss the main hurdle with defining an HKLL unitary, and point out a simple workaround.

Within fixed background QFT on the metric $g$, the unitary $\mathfrak{u}_s (\psi,a^{\mathcal{C}})$ is spacelike separated from $\phi_{p}$ and therefore commutes with it.
However, we are working not in fixed background QFT but in semi-classical gravity; here there is an additional complication.
Consider an HKLL reconstruction of a bulk operator of the form $O_{1} (t_{1})\dots O_{n} (t_{n})$.
We can think of each $O_i(t_i)$ as acting on a given bulk Cauchy slice by bulk Heisenberg evolution; in this case when we act this product of operators on a specific boundary state, 
$\ket{\psi}$, $O_{n}$ is Heisenberg-evolved using the equations of motion on the metric $g_\psi$, but $O_{i}$ is evolved using the equations of motion on a metric that includes the backreaction of $O_{i+1}\dots O_{n}$. Thus, the operator in the bulk is somewhat different from the intended operator.
We call this effect gravitational `spreading,' and it complicates the story. In particular, one of the $O_i$'s may in fact be inserted to the future of the peninsula in the state $O_{i+1}...O_n\ket{\psi}$. Thus, it is no longer the case that the HKLL operator $\mathfrak{u}_{s} (\psi,a)$ is guaranteed to commute with $\phi_{p}$ any more, \emph{even at leading order}.
Finally, because of entanglement wedge reconstruction, we expect the spreading to remain confined to the entanglement wedge of $A$.\footnote{In fact, because of the Python's lunch considerations discussed in Section \ref{sec:discussion} and the arguments in \cite{Engelhardt:2014gca}, we expect it to be confined to the outermost wedge.}
In order for $O_i$ to be to the past/future of the peninsula, it must have been inserted sufficiently early/late in boundary time $t_i$ --- times of order the (modular) scrambling time. 

One way to get around this issue then is to make sure the product of boundary operators only has support on a smaller domain of dependence. A simple solution is to consider a smaller boundary region $\tilde{A} \subsetneq A$, such that the causal surfaces of $A,\tilde{A}$ lie an $\mO(cG_{N})$ distance apart in the bulk, and $\mathcal{W}_{O}[\tilde{A}] \subset \mathcal{W}_{C}[A]$.
We expect such a region to exist, since we have already assumed that the state is smooth and so the size of the peninsula for $\tilde{A}$ should be similar to that for $A$.
By following the above construction for this smaller region, we get the same energy distribution as above up to $\mO (cG_{N})$ corrections --- both in position of the shocks as well as magnitude --- while at the same time ensuring that it commutes with $\phi_{p}$.\footnote{A potential confusion with this construction is that it allows the boundary to see outside the entanglement wedge of $\tilde{A}$. This is resolved by remembering that the signal reaches $A$ but not $\tilde{A}$ on the boundary.}

There are two points to note in conclusion.
First, the construction we gave here is not unique and may well not be the best one that both gives the right energy distribution and commutes with the signal.
Secondly, the \emph{exact} energy distribution in Section \ref{ssec:calculation} is in fact a solution of the Einstein equations which does not violate any known energy conditions; the only problem with it is that we do not know how to causally construct that solution by an operator acting on $A$.

%%%%%%%%%%%%%%%%%%%%%%%%%%%%%%%%%%%%%%%%%%%%%%%%%%%%%%%%%%%%%%
\section{Connection Between Bulk and Boundary Cocycle Flow} \label{sec:bd}

Recent developments in bulk reconstruction \cite{Faulkner:2017vdd,Chen:2019iro} have demonstrated that modular flow will be a key tool in decoding the interior of the black hole.
In this work, we have tried to argue for a \emph{bulk} semi-classical construction whose back-reaction allows us access to a peninsula region. This is to be contrasted with discussions of entanglement wedge reconstruction where the main ingredient is \emph{boundary} modular flow, which by \cite{Jafferis:2015del} is dual to modular flow in the entire entanglement wedge.
In this section, we show that in the cases considered in this paper the backreaction of the bulk Connes' cocycle flow has the same effect as boundary Connes' cocycle flow to leading order. Note that there are cases where the outermost quantum extremal surface can be far from the minimal entropy quantum extremal surface. In that case, boundary modular flow, which is dual to modular flow in the wedge bounded by the \emph{minimal} extremal surface, will be very different from the flow we describe here. In other words, the geometry resulting from boundary CC flow is only close to that resulting from the causal CC flow described in Sec. \ref{sec:gen} when the minimal QES is close to the causal surface.

Boundary CC flow is the CC flow with respect to the boundary algebra, and is directly related to the discussion of \cite{Faulkner:2017vdd,Chen:2019iro} in a simple way.
To see this, we follow the lead of \cite{MSY-2} and define a QFT correlator that formalises the ability to see beyond the causal wedge, that we shall call the `seeing correlator' $C_{see}$.
A boundary region $A$ with algebra $\mA_A$ has an entanglement wedge $\mW_E[A]$ whose algebra we will denote as $\mA_{E}$.
As in the previous sections, we take $A$ to be a spherical cap or union of disjoint spherical caps, so that there exists a base state $\ket{S_{\Omega}}$ whose boundary modular flow is local,
\begin{equation}
  \Delta_{S_{\Omega}; \mA}^{is} O(x) \Delta_{S_{\Omega}; \mA}^{-is} \propto O(x_{s}),
  \label{eqn:loc-mod-flow-7.0}
\end{equation}
where the proportionality factor is also local and depends on the spin and conformal weight of $O$. Here we are using the notation of Appendix \ref{app:aqft}, where $\Delta_{\phi;\mA}^{is}$ is the full modular flow unitary, which in density matrix notation is
\begin{align}\label{eqn:fullmodflow}
    \Delta_{\phi;\mA}^{is} = \rho_{\mA}^{is}\otimes \rho_{\mA'}^{-is}.
\end{align}

Similarly to Section \ref{sec:gen}, we will be interested in another state $\psi$ in which  $\mW_E[A]$ is larger than the causal wedge by an $\mathcal{O}(c G_{N})$ amount with $1 \gg c G_{N} \gg G_{N}$.\footnote{Note that in the previous section we only needed this constraint on $\mW_O[A]$. The current discussion in fact needs this to be true for the wedge bounded by the minimal $S_{gen}$ quantum extremal surface.} Given this constraint, the CC flow of interest is then the one with respect to the \emph{boundary} algebra $\mA$, $u_s(\psi,A) = \bigotimes_r \sigma_{A_r}^{is} \left(\rho^{\psi}_A\right)^{-is}$.

For an operator $\phi \in \mA_{E}$ that is not in the causal wedge and a local boundary operator $O \in \mA$, we know from bulk causality that $O$ cannot detect a $\phi$ insertion to leading order in $G_{N}$,\footnote{We actually require a more stringent restriction on this operator; it must create an excitation that is highly boosted and either its $T_{++}$ or its $T_{--}$ must vanish. We will take the outgoing (toward the boundary) case, where $T_{++} = 0$, in which case we find that $\phi$ can be brought into the past, and not the future, of $D(A)$. If $\phi$ has $T_-- = 0$ instead, we can only bring it into the future of $D(A)$.}
\begin{equation}
  \forall O \in \mA, \quad \mel**{\psi}{e^{-i \epsilon \phi} O e^{i \epsilon \phi}}{\psi} \approx \mel{\psi}{O}{\psi}.
  \label{eqn:zero-comm}
\end{equation}
Our claim is that $O$ can, however, detect an insertion of $\phi$ if it is followed by an application of the boundary CC flow. In other words, we make the replacement $e^{i \epsilon \phi} \to u_s(\psi,A) e^{i \epsilon \phi}$.
To first order in $\epsilon$ this detection is the same as the following commutator being $O(1)$,
\begin{align}
  \exists O \in \mA, \quad C_{see} =\mel**{\psi}{\left[ u_s(\psi,A)^{\dagger}\  O\  u_s(\psi,A), \phi \right]}{\psi} \sim G_{N}^{0}.
  \label{eqn:c-see}
\end{align}
This is of course not true for all $O \in \mA$.

Using the definition of $u_s(\psi,A)$ in \eqref{eqn:cocycle} and converting to the notation of equation \eqref{eqn:fullmodflow}, we have
\begin{equation}
  u_s(\psi,A)^{\dagger}\, O\, u_s(\psi,A) = \Delta_{\psi; \mA}^{is} \Delta_{\Omega; \mA}^{-is} O \Delta_{\Omega; \mA}^{is} \Delta_{\psi; \mA}^{-is}
  \label{eqn:cc-flow-split-id}
\end{equation}
and the replacement $\Delta_{\Omega; \mA}^{-is} O \Delta_{\Omega; \mA}^{is} \to \tilde{O}$, we find that
\begin{equation}\label{eqn:Cseemodflow}
  C_{see} = \mel**{\psi}{ \left[\tilde{O}, \Delta_{\psi;\mA}^{-is} \phi \Delta_{\psi; \mA}^{is} \right] }{\psi}.
\end{equation}
Thus, flowing with the cocycle amounts to studying boundary modular flow as in \cite{Faulkner:2017vdd}. The fact that the correlator $C_{see}$ in \eqref{eqn:Cseemodflow} is $\mathcal{O}(G_{N}^{0})$ instead of $\mathcal{O}(G_N)$ for some $\tilde{O}$ was a key assumption in the work of \cite{Faulkner:2017vdd}, which used boundary modular flow to implement entanglement wedge reconstruction.

In Sections \ref{ssec:peninsula-bd}, \ref{ssec:shock-bd} and \ref{ssec:gen-bd} we deal with three cases, where \ref{ssec:gen-bd} is the most general. In each of these cases we show that the seeing correlator becomes non-zero for precisely the same value of $s$ that brings the location of the operator $\phi$ into view in the corresponding bulk construction.
We will also be a little careful about defining the operator $\phi$ in these sections.

\subsection{``Seeing'' the Peninsula With Boundary Flow} \label{ssec:peninsula-bd}
In this subsection, we return to the set-up of Section \ref{sec:peninsula}, where we have two disconnected boundary regions regions $\mA_1$ and $\mA_2$, where $\mA_1$ is the right bath plus quantum mechanics system and $\mA_2$ is a subsystem of the left bath. The authors of \cite{AMM} noted that given access to both $\mA_1$ and $\mA_2$, one can actually send an excitation from the left black hole to the right via the action of a two-sided unitary of the form 
\begin{align}
U_{12}(g) = e^{i g \int f(x_1,x_2) O_1(x_1) O_2(x_2)}
\end{align}
where $f(x_1,x_2)$ is some smearing over the two regions $\mA_1$ and $\mA_2$ that cannot be factorized.    This is just a slight generalization of the procedure first laid out in \cite{GJW}. Here we will show that to leading order the boundary cocycle flow has the same effect as our bulk prescription: it reveals the peninsula to right observers.

We thus want to study states of the form
\begin{align}
\ket{\Omega_s} = u_s (\Omega,A) \ket{\Omega}
\end{align}
where we note again that this cocycle is defined on the boundary not in the bulk. Note that in the analogy to the work of Gao, Jafferis \& Wall, what we normally call ``modular time'' is playing the role of a coupling \cite{AMM}.
A similar observation appeared recently in \cite{Jafferis:2020ora}.

We now compute the seeing correlator as in equation \eqref{eqn:c-see}. If we define our ``message'' operator $\phi$ as being sent in from some time in the past of the left black hole, then the seeing correlator amounts to the response of a right-sided operator, $\phi_R$, to the insertion of a left-sided operator, $\phi_L$ when we deform the Hamiltonian by insertion of this unitary. In other words, we would like to compute the following response function
\begin{align}
\braket{e^{i \epsilon \phi_L(t_L)} u_s^{\dagger} \phi_R(t_R) u_s e^{-i\epsilon \phi_L(t_L)}}_{\Omega} = -i\epsilon \braket{[u_s^{\dagger} \phi_R(t_R) u_s,\phi_L(t_L)]}_{\Omega} + \mO(\epsilon^2)
\end{align}

Focusing on one of the terms in this commutator, we can use the standard formulae for the cocycle, as described in Section \ref{sec:splitflow} to re-write this as 
\begin{align}
\braket{\phi_R(t_R) \Delta_{S_{\Omega}; \mA_{12}}^{is} \Delta_{\Omega; \mA_{12}}^{-is} \phi_L(t_L)}_{\Omega} =\braket{\phi_R(t_R) \Delta_{\Omega; \mA_1}^{is} \Delta_{\Omega; \mA_{12}}^{-is} \phi_L(t_L)}_{\Omega} .
\end{align}
Interestingly, this correlator is very closely related to the one considered in the work of \cite{Balakrishnan:2017aa} which was used to prove the quantum null energy condition. Our work thus clarifies the role of shockwaves in the QNEC correlator: Connes' cocycle creates negative stress energy shocks that are probed by the operators $\phi_R,\,\phi_L$.

Now in order to compute this correlator, we would like to use the equivalence of bulk and boundary modular Hamiltonians \cite{Jafferis:2015del}, which states that we can switch boundary modular flow to bulk modular flow. Thus, we can write the correlator as
\begin{align}
\braket{\phi_R(t_R) \Delta_{\Omega; \mA_1}^{is} \Delta_{\Omega; \mA_{12}}^{-is} \phi_L(t_L)}_{\Omega} =\braket{\phi_R(t_R) \Delta_{\Omega; a_1}^{is} \Delta_{\Omega; a_{12}}^{-is} \phi_L(t_L)}_{\Omega}.
\end{align}
 The hard part will be to compute the action of $\Delta_{a_{12}; \Omega}^{is}$ on $\phi_L(t_L)$. Since this modular flow is associated to the disjoint union of two regions, it is non-universal. In the kinematics where $t_L$ becomes negative and large ($\geq$ the scrambling time), we can hope to use the simplification that its wave-function on the $t=0$ slice is largely supported near the quantum extremal surface of $a_{12}$. Since the size of the peninsula is of order $c\beta/\phi_r$, to leading order in this parameter we can use the fact that such excitations are effectively just boosted by the modular flow, with corrections to this coming at higher-orders in the distance to the RT surface, which we will assume is also $c\beta/\phi_r$.

To see this more explicitly, we can expand the correlation function in terms of single particle eigenstates on the $w^- = 0$ plane
\begin{align}
    \braket{\phi_R(t_R) \Delta_{a_1}^{is} \Delta_{a_{12}}^{-is} \phi_L(t_L)} = \int dw^-_L dw^-_R \braket{\phi_R|w^-_R} \bra{w^-_R} \Delta_{a_1}^{is} \Delta_{a_{12}}^{-is} \ket{w^-_L}\braket{w_L^-|\phi_L}.
\end{align}
Here we are using the Kruskal coordinates introduced in Section \ref{sec:peninsula}. For early enough $t_L$ and late enough $t_R$, these integrals will be dominated by the region of small $w_L^-$. In that region, $\Delta_{a_{12}}^{-is}$ acts as a simple boost about the quantum extremal surface, which we take to lie at $w^- = W_2^- = -w^+$. We can thus write 
\begin{align}
\Delta_{a_1}^{is} \Delta_{a_{12}}^{-is} \ket{w_L^-} \approx \Delta_{a_1}^{is} \ket{W_2^- + (w_L^--W_2^-)e^{2\pi s}} = \ket{w_L^- + (e^{-2\pi s}-1) W_2^-}
\end{align}
which is just a simple shift in the $w^-$ direction by an amount $W_2^-(e^{-2\pi s}-1)$. Thus, the correlation function becomes
\begin{align}
\braket{\phi_R e^{i P_- a^-} \phi_L}
\end{align}
where $a^- = W_2^-(e^{-2\pi s}-1)$. At large, positive $s$, this shift will bring any excitation between $w_L^- =0$ and $W_2^-$ into view of the the right side. This is precisely what we found happens in the bulk construction of Section \ref{sec:peninsula}.

\subsection{Boundary CC Flow in AdS-Vaidya} \label{ssec:shock-bd}
We now consider an AdS-Vaidya spacetime in which the gap between the entanglement and causal wedges is entirely created by a local unitary that creates a shock.
The reconstruction of operators in the entanglement wedge advocated for in \cite{AAL} is to merely conjugate the HKLL reconstruction in the shock-free geometry by the shock-creating unitary.
We show how the unitary CC flow ($s \in \mathbb{R}$) allows us to see an operator in the peninsula region.

The state is $U\ket{\beta}$, where $\ket{\beta}$ is a thermofield double of inverse temperature $\beta$ dual to a black hole of mass $M$ and $U = \mathds{1}_L \otimes U_R$ is a one-sided unitary that creates a shock of energy $\frac{E}{M} = (1+2 \alpha)^2 - 1$.
The metric and dilaton of a $1+1$-d eternal black hole are \eqref{eqn:bh-metric},\eqref{eqn:bh-dilaton}.
The bulk dual is of the usual AdS-Vaidya form, where the outside of the shock is a black hole metric of mass $M+E$, with coordinates $\tilde{w}^\pm$, and the inside metric is that of a black hole with mass $M$, with coordinates $w^\pm$.
The coordinate patches are patched satisfying two conditions, continuity of the radius/dilaton (physical condition) and the Schwarzchild time (gauge choice) \cite{Shenker:2013pqa}.
The set of points visible from the right boundary, which has $\tilde{w}^{-} < 0, \tilde{w}^{+} > 0$, is given by
\begin{align}
  \tilde{w}^{-} < 0\quad &\Rightarrow\quad w^{-} < -\frac{\alpha}{1-\alpha} e^{-\frac{2\pi}{\beta} t_{s}} \nonumber\\
  \tilde{w}^{+} > 0\quad &\Rightarrow\quad w^{+} > \frac{\alpha}{1-\alpha} e^{\frac{2\pi}{\beta} t_{s}}.
  \label{eqn:visible-pts}
\end{align}
We will need a one-parameter generalisation of these states, $U(t_s) \ket{\beta}$, where the shock has been evolved to time $t_s$.

We define our bulk operator as $\phi$ by the equation
\begin{equation}
  U(t_{s}) O(T) \ket{\beta} = \phi (t(t_{s},T), \lambda(t_{s},T)) U(t_{s}) \ket{\beta} + O(G_{N}) \equiv \phi (t_{s}, T) U(t_{s}) \ket{\beta} + O(G_{N}).
  \label{eqn:O-U-comm}
\end{equation}
Here, $O(T)$ is a boundary operator and $\phi$ is dressed to the RT surface, so that it is entirely spacelike to $U$ (for $t_s$ not too large).
We will consider an operator $\phi (0,T)$ for some $T$, so that $\phi$ is localised in the peninsula region.

Now that we have defined the bulk operator, we can ask about seeing and reconstructing it.
As above, our answer will be to use CC flow, but this time with the base state being the unperturbed TFD.
The CC flow operator just boosts the shock insertion unitary
\begin{align}
  u_s (U \beta, R) \ket{U O(T) \beta} = \mathds{1}_{L} \otimes \rho_{\beta}^{-is}  \rho_{U\beta}^{is} \ket{U O(T)\beta} = \rho_{\beta}^{-is} U \rho_{\beta}^{is} U^{\dagger} \ket{U O(T) \beta} = U(\beta s) O(T) \ket{\beta}.
  \label{eqn:cc-shock}
\end{align}

The action on the state is simply to move the shock in time while leaving the operator $O(T)$ untouched. The position of the horizon in the flowed state is given by \eqref{eqn:visible-pts},
\begin{equation}
    w_{c}^{\pm} = \pm \frac{\alpha}{1-\alpha} e^{\pm 2\pi s}.
\end{equation}
So, we see that the null separation between the horizon and the bifurcation surface decreases exponentially.

Note that on its own just relabeling the time coordinate does not in any meaningful way let you see more of the region behind the horizon. The cocycle $(D U \beta: D\beta)_s$ delays the shock, thus reducing the total center-of-mass energy between $U$ and $O$. This allows the $\phi$-excitation to reach the boundary.

\subsection{Seeing the Peninsula in the General Perturbative Case} \label{ssec:gen-bd}
The results of the previous two subsections are in fact just special cases of the more general fact that boundary modular flow has the same effect as bulk modular flow in the general perturbative case of section \ref{sec:gen}, when the gap between the HRT and causal surfaces is small but still parametrically larger than metric fluctuations.
The argument is fairly simple, and uses only elementary facts about modular flow.

To remind the reader, the general perturbative setup is as follows.
We have a boundary CFT${}_{d}$ dual to a bulk AdS${}_{d+1}$.
The region of interest $A \subset CFT_{d}$, with algebra $\mA$, is a ball or union of balls so that in a split vacuum $\ket{S_\Omega}$ its entanglement and causal wedges coincide and the modular flow is local in the bulk and boundary.
In the excited state $\ket{\psi}$, the EW and the CW do not coincide but the size of the gap between the HRT surface and the causal surface $\mathcal{C}$ scales with $cG_{N}$, where $1 \gg cG_{N} \gg G_{N}$.
In this section, we will use $X$ to denote boundary coordinates and $x$ to denote bulk coodinates.

With these assumptions we show that a bulk operator in the peninsula region can be seen by a boundary operator within the domain of development of $A$ conjugated with the boundary CC flow.
We will probe this by looking at the commutator
\begin{equation}
  C_{see,bd} (X,x) = \mel{\psi}{\left[ u_s (\psi,A)^{\dagger} O(X) u_s (\psi,A), \phi(x) \right]}{\psi},
  \label{eqn:c-see-bd}
\end{equation}
where $\phi$ is an operator in the peninsula region and $O$ is an operator on the boundary.
We will show that this commutator becomes nonzero with a sign corresponding to $O$ being in the future (past) of $\phi$ for $s$ greater than (lesser than) a critical value that is related to the distance of the $\phi$ operator from the CW and that tends to $\infty$ ($-\infty$) as it approaches the future (past) horizon of the EW.
For the purposes of this section, we will fix the position of $\phi$ by dressing it to the RT surface; different dressings shouldn't adversely affect the results of this section as long as they are localised to $EW[A]$.

To make the statements in the previous paragraph precise, let us erect a convenient coordinate system that will also aid us in calculating the commutator.
We denote boundary coordinates as $X = (t, \lambda^{a})$, defined so that for an operator $O \in \mA_r$
\begin{equation}
  \Delta_{S_\Omega;\mA}^{is} O(t,\lambda) \Delta_{S_\Omega;\mA}^{ -is} = \Delta_{\Omega;\mA_r}^{is} O(t,\lambda) \Delta_{\Omega;\mA_r}^{ -is} \propto O(t + 2\pi s, \lambda) \equiv O(X_{s}),
  \label{eqn:bd-mod-flow}
\end{equation}
where the proportionality factors are possible conformal and spin transformation weights.
Similar to Section \ref{sec:cc}, which component of $A$ $O$ belongs to is encoded in the parameters $\lambda$.

Now, we define bulk null coordinates $x^{\pm}$ that are affine parameters along their respective null rays.
We define them so that increasing $x^{+}$ and decreasing $x^{-}$ go towards the boundary, and that $x^{\pm}$ are functions of the boundary point they intersect.
Since we are only interested in a small region outside the causal wedge and the geometry of the bulk differs from the geometry dual to the base state only perturbatively, we expect that this parameterization can be continuously extended to the entire entanglement wedge.
Finally, we fix additive and multiplicative ambiguities in the parametrisations by the following two conditions,
\begin{align}
  x^{\pm}|_{\mathcal{C}} &= 0 \nonumber\\
  \Delta_{S_\Omega;\mathcal{A}}^{is} \phi (x^{+},x^{-},\lambda) \ket{S_\Omega} &\propto \phi (x^{+} e^{2\pi s}, x^{-} e^{-2\pi s}, \lambda) \ket{\Omega},
  \label{eqn:xpm-defn}
\end{align}
where $\mathcal{C}$ is the causal surface and the second equation defines the modular flow of the bulk operator \emph{in the base state geometry}.
Both of these are matters of convenience and only serve to lighten the notation.
These two statements also mean that
\begin{equation}
  x^{\pm} (X_{s}) = e^{\pm 2\pi s} x^{\pm} (X),
  \label{eqn:bd-mod-flow-to-bulk-null}
\end{equation}
where $X_{s}$ was defined in \eqref{eqn:bd-mod-flow}.

In these coordinates, the statement we prove is that for $s > 0$,
\begin{align}
  C_{see} (X, \delta x^{\mu}) \neq 0 \quad \Rightarrow \quad \delta x^- - \delta x_E^- < x^{-} (X) + e^{-2\pi s} \delta x_{E}^{-} < e^{-2\pi s} \delta x_E^-.
  \label{eqn:seeing-statement}
\end{align}
This agrees with what we get from the backreaction of the bulk modular flow.

The first step in the proof is to use the standard property of CC flow that
\begin{equation}
  u_s (\psi,A)^{\dagger} O(X) u_s (\psi,A) = \Delta_{\psi;\mA}^{is} \Delta_{S_\Omega;\mA}^{-is} O(X) \Delta_{S_\Omega;\mA}^{is} \Delta_{\psi;\mA}^{-is} \propto \Delta_{\psi}^{is} O(X_{-s}) \Delta_{\psi}^{-is}
  \label{eqn:cc-bd}
\end{equation}
Picking the term of the commutator with $\phi$ on the right for definiteness, the $\Delta_{\psi}$ on the left of the above expression is absorbed by the identity $\Delta_{\psi} \ket{\psi} = \ket{\psi}$, and the $\Delta_{\psi}$ on the right plays the role of modular flowing $\phi$.
Since boundary modular flow is the same as bulk modular flow in the EW and $\phi$ is in the peninsula, the modular flow is (to leading order in $G_N$)
\begin{equation}
  \Delta_{\psi;\mA}^{-is} \phi (\delta x^{-}) \ket{\psi} = \phi (\delta x_{E}^{-} + (\delta x^{-} - \delta x_{E}^{-}) e^{ 2\pi s}) \ket{\psi}.
  \label{eqn:bulk-mod-flow}
\end{equation}
One might be worried about the fact that since the perturbed HRT surface isn't as symmetric as the unperturbed one, there is no preferred point to boost about --- i.e. in writing this expression we have tacitly chosen that the boost is about the point at the same value of $\lambda^a$.\footnote{We expect that this can be calculated explicitly in Euclidean path integral states as in \cite{Faulkner:2016mzt,BP,BP2}.}
However, it can easily be checked that this choice contributes at higher orders in $\delta x_E^\pm - \delta x^{\pm} \sim c G_N$.

The bulk null separation between the modular flowed operators is
\begin{equation}
  x^{-} (X) e^{ 2\pi s} - \delta x_{E}^{-} - \left( \delta x_{E}^{-} - \delta x^{-} \right) e^{2\pi s}  = e^{2\pi s}  \left\{ x^{-} (X) - \delta x^{-} - \delta x_{E}^{-} (e^{- 2\pi s} - 1) \right\}.
  \label{eqn:bd-mod-flow-answer}
\end{equation}
Since the commutator vanishes unless the flowed points are null or time-like separated, we find \eqref{eqn:seeing-statement} as promised.

\section{Discussion} \label{sec:discussion}
We now briefly discuss some interesting connections with other work and possible future directions. 
\subsection*{Relation with the Python's Lunch Conjecture}
As discussed in the Introduction, quantum focusing prevents the ``causal operations'' discussed in this paper from causally revealing the region beyond $\mathcal{W}_{O}[A]$. This actually dovetails nicely with recent work on the complexity of bulk reconstruction. In \cite{PL}, the authors conjecture that when $\mathcal{W}_{E}[A]$ includes non-minimal quantum extremal surfaces, bulk reconstruction beyond the outermost quantum extremal surface is exponentially complex in $N$. 

More explicitly, the authors of \cite{PL} conjectured that, in the classical bulk limit, whenever there is a non-minimal area extremal surface there is also a so-called ``maximinimax'' surface, whose area $A_{max}$ is a maximum with respect to some foliation by co-dimension two surfaces of the Cauchy slice on which it lies. Intuitively, it is the cross sectional area of the ``bulge'' between two local minima. The authors then conjectured that the complexity of reconstruction goes like 
\begin{align}
\mathcal{C} \sim e^{\left(A_{\text{max}} - A_{\text{outermost}}\right)/8G}.
\end{align}

In other words, the complexity for reconstruction should be exponential in the number of UV degrees of freedom, $N^2$. The protocols we have outlined in this paper are enacted by applying a unitary in the low-energy bulk effective field theory, restricted to the causal wedge. Such operators are expected to be ``simple''. Thus, it is consistent with the Python's lunch conjecture that our protocol for seeing beyond the causal wedge fails to give the bulk observer access beyond the outermost RT surface.\footnote{We thank Geoff Penington for pointing this out. Further development of this connection and the extension of the seeing procedure to classical geometries with large gaps between $\mathcal{W}_{O}[A]$ and $\mathcal{W}_{C}[A]$ is the subject of \cite{Arvin:FutureWork22}.} 

Note, however, that action with the \emph{boundary} cocycle flow does have the requisite exponential complexity. As discussed in the previous section, boundary cocycle flow is directly related to boundary modular flow, which has been shown to bring operators out from entanglement islands \cite{Chen:2019iro,Faulkner:2017vdd}. In such situations, the boundary modular flow is expected to have exponential complexity by the results of \cite{Bouland:2019pvu} and thus this is consistent with the Python's lunch conjecture.

\subsection*{Bulk Reconstruction vs. Seeing}

Let $R$ be a codimension zero region of a fixed spacetime where a QFT lives. It was proposed in \cite{haag1962postulates} that the algebra of local QFT operators associated to R is in fact equivalent to the algebra associated to the maximal causal completion of this region (which includes all points $p$ that are both in the past and the future of $R$). Though this statement is not proven in its most general form, there are explicit setups where it can be shown to hold for all reasonable QFTs \cite{borcherstt,arakitt}.\footnote{See \cite{Witten:2018zxz} for a nice review.}

In our AdS/CFT setup, depending on the sign of $s$, we find two \emph{different} geometries in which $\mathcal{P}[A]$ is in the future and past of $D(A)$ respectively. It seems plausible that an extension of the above statement to QFT on different geometries could connect our results to bulk reconstruction of operators in $\mathcal{P}[A]$: that we can rewrite operators localized to $\mathcal{P}[A]$ as a combination of near boundary bulk operators in $D(A)$.

\section*{Acknowledgements}
We thank Ahmed Almheiri, Raphael Bousso, Netta Engelhardt, Victor Gorbenko, Nirmalya Kajuri, Nima Lashkari, Roberto Longo, Raghu Mahajan, Juan Maldacena, Onkar Parrikar, Geoff Penington, Karl-Henning Rehren, Douglas Stanford, Alex Streicher, Edward Witten and Ying Zhao for helpful discussions. We also thank Raphael Bousso and Juan Maldacena for comments on the draft. ASM was supported by the Berkeley Center for Theoretical Physics; by the Department of Energy, Office of Science, Office of High Energy Physics under QuantISED Award DE-SC0019380 and under contract DE-AC02-05CH11231; and by the National Science Foundation under grant PHY1820912 ; and by the National Science Foundation under Award Number 2014215.
Parts of this work were done during the programs ``Workshop on Qubits and Spacetime'' at IAS, Princeton and ``Geometry from the Quantum'' at KITP, UCSB.

\appendix 
\section{Constructing $\psi_s$ for General Algebras} \label{app:aqft}
In this appendix, we discuss our field theoretic construction detailed in Section \ref{sec:cc} without assuming the existence of density matrices associated to regions. As we will see, our states $\ket{\psi_s}$ are well-defined directly in the continuum limit. Our construction follows closely the work of Ceyhan \& Faulkner \cite{CF}. We merely extend their calculations to include split vacua. We begin with a brief review of the main ingredient in our construction: the cocycle.

\subsection{Review of Connes' Cocycle}
Given a spacetime region $A$ and an associated algebra $\mA_A$ which acts on the Hilbert space of the theory $\mH$, the relative conjugation operator $S_{\psi | \phi}$ between two states $\psi,\,\phi$ is defined to act as 
\begin{align}
S_{\psi | \phi; \mA_A} \alpha \ket{\psi} = \alpha^{\dagger} \ket{\phi}
\end{align}
for all $\alpha \in \mA_A$.\footnote{We will assume  that $\psi$ and $\phi$ are both cyclic and separating to lighten the notational load, although this assumption can be relaxed \cite{CF}.} For the remainder of this section, unless necessary, we will drop the subscript on the conjugation operator that labels which algebra it acts with respect to. Using the polar decomposition of the conjugation operator, we can write $S_{\psi | \phi} = J_{\psi | \phi} \Delta_{\psi | \phi}^{1/2}$ where $J_{\psi | \phi}$ is an anti-unitary operator and $\Delta_{\psi | \phi}$ is called the \emph{relative modular operator}. From the polar decomposition, we can write
\begin{align}
S_{\psi | \phi}^{\dagger} S_{\psi | \phi} = \Delta_{\psi | \phi}.
\end{align}

In terms of density matrices, the modular operator takes the form
\begin{align}
\Delta_{\psi | \phi; \mA_A} = \rho_{\psi;\mA_A'} \otimes \rho^{-1}_{\phi;\mA_A}.
\end{align}
A nice object to consider is the product of two modular operators which is called the \emph{Connes cocycle}
\begin{align}\label{eqn:cocycle-a}
(D\phi : D\psi)_s \equiv \Delta_{k|\phi}^{is} \Delta_{k|\psi}^{-is}=\Delta_{\phi}^{is}\Delta_{\phi | \psi}^{-is} = \Delta_{\psi | \phi}^{is} \Delta_{\psi}^{-is}
\end{align}
where we have used the notation $\Delta_{\phi} \equiv \Delta_{\phi | \phi}$. Here we have used the non-trivial fact that this product of relative modular operators is independent of the state $k$ \cite{CF,Witten:2018zxz}. Two convenient choices for $k$ are then $\phi$ and $\psi$, as shown in \eqref{eqn:cocycle-a}. For algebras that admit density matrices, the cocycle takes the form
\begin{align}
\mathbf{1}_{\mA_A'} \otimes \rho_{\phi;\mA_A}^{is}\rho_{\psi;\mA_A}^{-is} \in \mA_A
\label{eqn:cc-rhos}
\end{align}
which is a unitary entirely in the algebra $\mA_A$. Indeed, this property persists in the continuum and it can be shown that $(D\phi:D\psi)_s \in \mA_A$ for all real $s$ using Tomita-Takesaki theory \cite{CF, Witten:2018zxz}.

We will now follow the lead of \cite{CF} and consider the family of states
\begin{align}
(D\phi : D\psi)_s \ket{\psi} \equiv \ket{\psi_{\phi}^s}.
\end{align}
This family has the property that expectation values of observables in $\mA_A'$ are untouched, but observables in $\mA_A$ are flowed with respect to the $\phi$ state. This can be seen by noting that 
\begin{align}
(D\phi: D\psi)_s \ket{\psi} = \Delta_{\psi | \phi; \mA_A}^{is} \ket{\psi} = \Delta_{\phi| \psi;\mA_A'}^{-is} \ket{\psi} = \Delta_{\phi;\mA'}^{-is} (D\phi: D\psi; \mA_A')_s\ket{\psi}.
\end{align}
In the various equalities in this formula, we have used the simple facts that $\Delta_{\psi}^{is} \ket{\psi} = \ket{\psi}$, $\Delta_{\psi| \phi;\mA_A} = \Delta_{\phi | \psi;\mA_A'}^{-1}$ and the definition of the cocycle, respectively. The first two of these facts can be verified easily under the assumption that density matrices for $\mA_A$ and $\mA_A'$ exist. For the derivation of these facts for general von Neumann algebras, see \cite{CF, Witten:2018zxz}.

The upshot is that the cocycle for $\mA_A$ acting on $\ket{\psi}$ produces a state which, up to a unitary on $\mA_A'$, looks like it has been flowed by the modular operator of the reference state $\phi$. Using the non-trivial fact that modular flow preserves the algebra (i.e. $\Delta_{\phi;\mA_A}^{-is} \mA_A \Delta_{\phi; \mA_A}^{is}=\mA_A$), we have that 
\begin{align}\label{eqn:expvalueflow}
\braket{\mO_{\mA_A}}_{\psi_{\phi}^s} = \bra{\psi} \Delta_{\phi;\mA_A}^{-is} \mO_{\mA_A} \Delta_{\phi; \mA_A}^{is}\ket{\psi}.
\end{align}
This will be the key fact that we will need in the following construction. Note that when $\phi$ is the vacuum and $A$ has a conformal killing vector, then \eqref{eqn:expvalueflow} becomes \eqref{eq-boost-transform} in the main text.

\subsection{Cocycle With Respect to Split States in Quantum Field Theory} \label{sec:splitflow}
We now consider a specific cocycle on an algebra which is given by the algebraic union of $n$ algebras $\mathcal{A}_1, \cdots,  \mathcal{A}_n$ belonging to the disconnected regions $A_{1}, \cdots, A_{n}$ described in subsection \ref{sub-summary}. In this Appendix, we will mostly work explicitly with the case where $A_{r}$ is a Rindler wedge region in Mink$_{1+1}$ and $n=2$ or AdS$_{d+1}$ and $n\geq 2$. We could also consider spherical regions in Minkowski  with $d>1$ where the theory is a CFT. In that case, many of the formulae listed below need to be supplemented with the relevant conformal factors. We leave the description of that case as an excercise for the reader.

The cocycle of interest will be between a general state, $\ket{\psi}$, and a special state $\ket{S_{\Omega}}$ which has no correlations between different $\mA_r$ but agrees with the vacuum state on each algebra individually. We will call such states \emph{split vacuum states}. In the case that $\psi = \Omega$ is the vacuum and our theory is on pure AdS, we show in Sec. \ref{sec:peninsula} that the state
\begin{align}
(DS_{\Omega} : D\Omega)_s \ket{\Omega} = \ket{\Omega_s}
\end{align}
has the right energetic properties to move the causal wedge up to the quantum extremal surface.

\subsubsection*{Split States}
We will now define the split state more rigorously. In finite dimensional quantum systems, split states are very easy to construct. Given a density matrix $\rho_{A}$ on $\bigotimes_r \mA_r$, then a split state is just a purification of the density matrix $\sigma_{A}$ given by 
\begin{align}
\sigma_{A} =\bigotimes_r \rho_r
\end{align}
where $\rho_r$ is the reduced density matrix associated to $\rho_A$ on one of the component algebras $\mA_r$. 

In a continuum quantum field theory, there are several equivalent ways of formalizing split states. The definition that will be convenient for us is the following. If we have two algebras $\mA_{1,2}$ such that $\mA_1 \subset \mA_2'$, then the algebra $\mA_1 \vee \mA_2$ obeys the split property if and only if there exists a unitary
\begin{align}
U: \mH \to \mH \otimes \mH,
\end{align}
 which maps the Hilbert space of the theory to a tensor product with itself $n$-times. Furthermore, the unitary must satisfy the equations
 \begin{align}
 U \mA_1 U^{-1} = \mA_1 \otimes \mathbf{1},\  U \mA_1 \vee \mA_2 U^{-1} = \mA_1 \otimes \mA_2 
 \end{align}
 It is important that $U$ is a \emph{unitary} and not an isometry. That such a unitary exists relies on the continuum nature of the theory.\footnote{Note that in the doubled Hilbert space, there is an obvious type I factor, namely just the space of bounded operators $\mathcal{B}(\mH)$ on the first tensor product factor. If we define $\mathcal{N} \equiv U^{-1} \left(\mathcal{B}(\mH) \otimes \mathbf{1} \right) U$ then we have the nesting property $\mA_1 \subset \mathcal{N} \subset \mA_2'$. In fact, one can show that the existence of such a nested  type-I factor is equivalent to the existence of the unitary $U$ with properties in the main text \cite{DF}.}
 
We then define a split state $\ket{S_{\psi}}$ as 
\begin{align}
\ket{S_{\psi}} = U^{-1}\left( \ket{\psi} \otimes \ket{\psi}\right).
\end{align}
Using this unitary, one can check that indeed, 
\begin{align}
\bra{S_{\psi}} \mO_1 \mO_2 \ket{S_{\psi}} = \bra{S_{\psi}} \mO_1 \ket{S_{\psi}}\bra{S_{\psi}} \mO_2 \ket{S_{\psi}} = \bra{\psi} \mO_1 \ket{\psi}\bra{\psi} \mO_2 \ket{\psi}
\end{align}
for $\mO_{1,2} \in \mA_{1,2}$. Furthermore, the image of the vacuum under this unitary, which we call $\ket{\xi} = U\ket{\Omega}$, is an entangled state of two copies of the original field theory. 

The generalization to split states on multiple non-overlapping regions is the natural one: for $n$ disconnected regions, the unitary $U$ will map from a single copy of $\mH$ to a an $n$-fold tensor product of $\mH$. The $r$'th algebra will then be mapped to an algebra which acts on the $r$'th tensor product factor.

Note that split states have infinite energy when any two of the $A_r$'s have touching boundaries. When there is a gap between all of the component regions, it has been argued in a large class of examples that such split states do indeed exist \cite{Fewster:2016aa, Doplicher:aa, Morinelli_2017}. For the purposes of our discussion, we will take as an axiom of our theory that split vacua exist on the algebra of interest, $\bigvee_r \mA_r$, unless otherwise stated.

An important side point is that split vacua do not strictly exist between two non-overlapping Minkowski-Rindler wedges in dimensions $d>1$. As detailed in \cite{Witten:2018zxz}, this is because the two Rindler wedges remain ``close'' even as we move off to infinity along the transverse directions. This leads to operators with bounded fluctuation in split-vacua but non-bounded fluctuations in the global vacuum. Thus, the algebra $\mA_1 \vee \mA_2$ and $\mA_1 \otimes \mA_2$ are no longer isomorphic, since $\mA_1 \vee \mA_2$ will not include operators with unbounded vacuum fluctuations.

This issue does not apply, however, to two non-overlapping AdS-Rindler wedges since their relative proper distance diverges as one moves towards the boundary. Note also that AdS-Rindler wedges are dual to boundary spheres. These spheres admit a boundary CFT split state in any boundary dimension $d \geq 1$ and so we also expect the bulk theory to admit a dual split state as well. In this Appendix, we thus work with Rindler wedges in Minkowski$_{1+1}$ space or AdS$_{d+1}$ with $d>1$.

Note there does not exist a unique split state. The unitary $U$ can be modified by the action of a unitary $V' \in \mA_A'$.\footnote{The authors in \cite{DF} were interested in a unique split state which also lies in the so-called natural self-dual cone of the vacuum. Such split states have the property that each tensor factor is invariant under CRT.} Fortunately, the properties of the state $\ket{\Omega_s}$ we are interested in will be independent of $V'$. We turn to a more detailed analysis of this state now.

\subsubsection*{Cocycle with respect to a split state}
From equation \eqref{eqn:expvalueflow}, we see that expectation values of observables in $\mA$ are just flowed by the split state's modular flow. Let's take an example of an operator in $\mA$ which is a product of local operators like $\mO_1 \mO_2$ where $\mO_{i} \in \mA_{i}$. From equation \eqref{eqn:expvalueflow}, we have
\begin{align}
\braket{\mO_1 \mO_2}_{\psi_s} = \bra{\psi} \Delta_{S_{\Omega};\mA}^{-is} \mO_1 \mO_2 \cdots \mO_{n} \Delta_{S_{\Omega};\mA}^{is}\ket{\psi}.
\end{align}
We can understand this correlation function simply upon mapping to the multi-copy Hilbert space and back. Using the fact that under the unitary $\mO_1 \mO_2 \to \mO_1 \otimes \mO_2$ and that $\Delta_{S_{\Omega}; \mA_{12}} \to \Delta_{\Omega;\mA_1}\otimes \Delta_{\Omega;\mA_2}$, we find the simple answer 

\begin{align}
\braket{\mO_1 \mO_2}_{\psi_s} = \bra{\psi}\left( \Delta_{\Omega;\mA_{1}}^{-is} \mO_1  \Delta_{\Omega;\mA_{1}}^{is}\right) \left( \Delta_{\Omega;\mA_{2}}^{-is}\mO_2 \Delta_{\Omega;\mA_2}^{is}\right)\ket{\psi}.
\end{align}

In other words, the operators get boosted in their respective wedges $D(A_{i})$. This generalizes in the obvious way to the case of multiple local operators within each algebra $\mA_{i}$. Note that this de-correlates the two regions and so in some sense brings the state closer to a split state as $s \to \infty$. In the language of \cite{Lashkari:2019ixo}, the cocycle has the effect of sewing a split vacuum into $\mA_1 \vee \mA_2$.\footnote{The authors of \cite{Lashkari:2019ixo} showed that in the $s \to \infty$ limit, this intuition actually becomes precise in the weak-topology.} We now turn to proving basic relationships between the relative entropy and the null energy in $\psi_s$. This will help justify our understanding of the energy distribution in $\ket{\psi_s}$ in the main text. 

\subsection{Deriving the Averaged Null Energy Distribution in $\ket{\psi_s}$}\label{sec:flowedANE}
We now derive an equation for the amount of null energy (ANE) in the flowed states using techniques that do not rely on Hilbert space factorization.
In this subsection, we will calculate what we call the \emph{split null momentum operator}, which we can think of as an abstract generalization of the averaged null energy to disconnected (AdS) Rindler wedges.

As above, we consider a region $A = \cup_r A_r$ which is the union of $n$ disconnected (AdS) Rindler wedges. We define the region $B = B_1 \cup \left(\bigcup_{r=2}^n A_r\right)$, where $B_1$ is defined as a sub-region of $A_1$ bounded by an entangling surface which is a null translation of $\partial A_1$ by an amount $\Delta x^{+}$. Following the notation of Section \ref{sec:cc}, around each component Rindler wedge we have set up null coordinates so that the $x^+/x^-$ coordinate increases toward the future and toward the interior/exterior of the wedge.
 
For a single Rindler wedge, the total averaged null energy can be defined using the algebra of half-sided modular inclusions for (AdS) Rindler wedges as \cite{CF,Longo:2017aa,Araki:aa}
\begin{align}\label{eqn:differentiateP}
P =\left. i\frac{d}{d\Delta x^+} \right \vert_{\Delta x^+=0} \left.\frac{d}{ds} \right \vert_{s=0} \left( \Delta_{\Omega;\mA_{A_1}}^{-is} \Delta_{\Omega;\mA_{B_1}}^{is}\right).
\end{align}
We define a new quantity called the \emph{split null momentum} operator, which we define analogously to the null momentum operator as
\begin{align}\label{eqn:differentiatePs}
P^1_+ =\left. i\frac{d}{d\Delta x^+} \right \vert_{\Delta x^+=0} \left.\frac{d}{ds} \right \vert_{s=0} \left( \Delta_{S_{\Omega};\mA_A}^{-is} \Delta_{S_{\Omega};\mA_B}^{is}\right).
\end{align}
In the multi-copy Hilbert space, $P_1$ takes the more familiar form of 
\begin{align}
    P_1 = U^{\dagger}(P \otimes \mathbf{1}...\otimes  \mathbf{1})U.
\end{align}
We denote by $P_r = U^{\dagger} (\mathbf{1}\otimes... P \otimes...\mathbf{1}) U$ the split null momentum for the $r$'th Rindler wedge, $A_r$.

We would like to understand the energy distribution in the doubled Hilbert space. We would now like to study
\begin{align}
\braket{P_1}_{\psi_s} - \braket{P_1}_{\psi} = \braket{P \otimes \mathbf{1}...\otimes  \mathbf{1}}_{U\psi_s} - \braket{P \otimes \mathbf{1}...\otimes  \mathbf{1}}_{U\psi}.
\end{align}

In the multi-copy Hilbert space, the cocycle just gets mapped to $U(DS_{\Omega}: D\psi; \mA_{1\cup 2})_s U^{\dagger} = \left(D \Omega^{\otimes n}:D U\psi; \bigotimes_r\mA_r\right)_s$. States very similar to 
\begin{align}
 \left(D \Omega^{\otimes n}:D U\psi; \bigotimes_r\mA_r\right)_s\,U\ket{\psi}
\end{align}
were studied in \cite{CF}. In that work, Ceyhan \& Faulkner were able to show that 
\begin{align}
\braket{P}_{\psi^{CF}_s} - \braket{P}_{\psi} = (1-e^{-2\pi s})\frac{d}{dX^+} S(\psi |\Omega;\mA).
\end{align}
where $\ket{\psi^{CF}_s} = (D\Omega: D\psi; \mA_1)_s \ket{\psi}$. The difference with our setup is just that Ceyhan \& Faulkner worked with states defined on a single copy of the Hilbert space. The important aspect for them was that the vacuum state modular flow obeys a half-sided modular inclusion algebra.\footnote{Note that also the authors of \cite{CF} worked with Rindler wedges in Mink$_{d+1}$ but could have been working with Rindler wedges in AdS$_{d+1}$ since the algebra of half-sided modular inclusions is present there as well.} Fortunately, the split vacuum modular flow also obeys an exactly analogous algebra on the multi-copy Hilbert space, which takes the form
\begin{align}
    \Delta_{\Omega^{\otimes n}; \mA_1 \otimes \mA}^{is} \Delta_{\Omega^{\otimes n}; \tilde{\mA}_1 \otimes \mA}^{-is} = \Delta_{\Omega; \mA_1}^{is} \Delta_{\Omega; \tilde{\mA}_1}^{-is}\otimes \mathbf{1} = e^{i(e^{2\pi s}-1) P\Delta x^+}\otimes \mathbf{1}
\end{align}
where $\tilde{\mA}_1 \subseteq \mA_1$ is a sub-algebra that has been null translated by an amount $\Delta x^+$ and $\mA = \bigotimes_{r=2}^{n} \mA_r$. This difference between the two set-ups should only affect the notational baggage of the calculations, but not the mechanics. Thus, one finds
\begin{align}\label{eqn:splitnullenergy}
\braket{P \otimes \mathbf{1}...\otimes  \mathbf{1}}_{U\psi_s} - \braket{P \otimes \mathbf{1}...\otimes  \mathbf{1}}_{U\psi} =  (1-e^{-2\pi s})\left( \frac{d}{dX^+}\otimes \mathbf{1} \right)S_{\text{rel}}\left(U\psi \,\left|\, \Omega^{\otimes n}\,;\ \bigotimes_r \mA_r\right.\right)
\end{align}
where by $\left(\frac{d}{dX^+}\otimes 1\right)$ we mean a null derivative with respect to the region whose algebra acts on the first Hilbert space factor. Formula \eqref{eqn:splitnullenergy} will be important for proving differentiability of the relative entropy in the next section.

\subsubsection*{Total ANE for $n=2$ in Mink$_{1+1}$}
As a side calculation, in the case when we have two non-overlapping Rindler wedges in Mink$_{1+1}$, we can actually derive a simple formula for the \emph{total} averaged null energy in equation \eqref{eqn:differentiateP}. We will find that the total null momentum in the $\psi_s$ state takes the form
\begin{align}\label{eqn:energyapp}
\braket{P}_{\psi_s} - \braket{P}_{\psi} = (1-e^{-2\pi s})\frac{d}{dX_2^+} S(\psi | S_{\Omega};\mA_{1\cup2})+ (e^{2\pi s}-1) \frac{d}{dX_1^+} S(\psi|\Omega;\mA_{1\cup 2}).
\end{align}
Here we are not using the funny coordinate system detailed above equation \eqref{eqn:differentiateP}, but rather the normal background coordinates for Mink$_{1+1}$ with metric $ds^2 = -dx^+ dx^-$. Note that by monotonicity of relative entropy, this quantity is positive for all $s$. 

Then, using equation \eqref{eqn:splitnullenergy} for the two split null energies $P_+ \otimes \mathbf{1}$ and $\mathbf{1}\otimes P_+$, we see that to prove \eqref{eqn:energyapp} we just need the relation
\begin{align}\label{eqn:splitANE}
\braket{P_+}_{\psi_s} - \braket{P_+}_{\psi} = \braket{P_+\otimes 1}_{U\psi_s} - \braket{P_+\otimes 1}_{U\psi} + \braket{1 \otimes P_-}_{U\psi_s} - \braket{1 \otimes P_-}_{U\psi}.
\end{align}

To prove \eqref{eqn:splitANE}, consider the unitary $V_t$ given by 
\begin{align}
V_t =U^{\dagger} e^{-i t \left(1\otimes P_+ + P_+\otimes 1\right)}U e^{itP}.
\end{align}
For any operator $\mO_2 \in \mA_2$ and for $t>0$, we have that
\begin{align}\label{eqn:invariance}
V_t^{\dagger} \mO_2 V_t = \mO_2
\end{align}
since the action of $U^{\dagger} e^{-i t \left(1\otimes P_+ + P_+\otimes 1\right)}U $ just does the following: maps $\mO_2$ to $1 \otimes \mO_2$, shifts it in the future $x^+$ direction by an amount $\Delta x^+ = t$ and then maps back to the single copy. This produces just $e^{itP_+} \mO_2 e^{-itP_+}$, which is then undone. Note that $t>0$ was important here; if $t<0$, then we are not guaranteed that $1\otimes \mO_2$ gets mapped within $1 \otimes \mA_2$. 

To extend formula \eqref{eqn:invariance} to an interval of $t<0$, we can take our split state to be split with respect to a union of non-overlapping Rindler wedges that are slightly larger than $A_1 \cup A_2$. Doing this guarantees that there is an open interval of $t<0$ for which the doubling unitary $U$ maps $e^{it(1\otimes P_+)} (1\otimes \mO_2) e^{-it(1\otimes P_+)}$ back to a local operator in $\mA_{1\cup 2}'$. Thus, we find that for such split states equation \eqref{eqn:invariance} holds for $t \in (-\epsilon,\infty)$ for some $\epsilon>0$.

The same manipulations show that for all $\mO_1 \in \mA_1$
\begin{align}
V_t^{\dagger} \mO_1 V_t = \mO_1
\end{align}
but now for $t \in (-\infty, +\epsilon)$ for some $\epsilon >0$.\footnote{We are using conservation of stress energy to move $P$ between the two algebras $\mA_1$ and $\mA_2$. This requires that $\braket{T_{+-}}$ go to zero at $x^+ \to \pm \infty$, which we take as an assumption.} The upshot is that for $t \in (-\epsilon , \epsilon)$ for some $\epsilon >0$, $V_t$ commutes with all operators in $\mA_1$ and $\mA_2$. $V_t$ then commutes with all operators in $\mA_1 \vee \mA_2$. In particular, it commutes with the cocycle $D(S_{\Omega}:\psi;\mA_1 \vee \mA_2)_s$.

Then we have that 
\begin{align}
\braket{V_t}_{\psi_s} = \braket{V_t}_{\psi}
\end{align}
for $t$ in some small interval about $0$. Differentiating this equation about $t=0$, we find equation \eqref{eqn:splitANE} and thus prove equation \eqref{eqn:energyapp}.

\subsection{Differentiability of the Relative Entropy for the $\psi_s$ State }
\label{app:sumrule}
In the main text, we assumed differentiability of the relative entropy in the $\psi_s$ state to derive formula \eqref{eq-energyresult}, which we reproduce here for convenience:
\begin{align}\label{eq-energyresultapp}
    \int_{0^-}^{\infty} dx^+_r \langle T_{++}(x^-_r=0, x^+_r, y^i_r)\rangle_{\psi_{s}} - \int_{0^+}^{\infty} dx^+_r \langle T_{++}(x^-_r=0, x^+_r, y^i_r)\rangle_{\psi_{s}} = \frac{e^{-2\pi s}-1}{2\pi \sqrt{H(y_r)}}\left.\frac{\delta S(\rho^{\psi}_{a(X^+_r)})}{\delta X^+_r(y^i_r)} \right|_{X^{+}_r=0},
\end{align}
where we use the notation of Sec. \ref{sec:cc}. In this sub-section, we will argue for differentiability of the relative entropies $S(\psi_s | S_{\Omega};\mA)$ with respect to the endpoints of $\mA$. Here we are working in general dimensions and where $\mA = \cup_r \mA_r$ is the union of a countable number of Rindler wedges. Note that for $n>2$, we must work in AdS$_{d+1}$ since one cannot construct more than two non-overlapping Rindler wedges in Minkowski.

To argue for differentiability, we follow exactly the steps of \cite{CF}, but in the doubled Hilbert space. One of the key formulae that \cite{CF} used to prove differentiability was the so-called ``sum-rule,'' which relates the averaged null energy to relative entropy variations. The sum rule on a single copy of the Hilbert space takes the form
\begin{align}\label{eqn:normalsumrule}
\frac{d}{dX^+} S(\psi | \Omega; \mA') - \frac{d}{dX^+} S(\psi | \Omega; \mA) = \int_{-\infty}^{+\infty} dx^+ \sqrt{H} \braket{T_{++}}_{\psi} \equiv \braket{P}_{\psi}
\end{align}
where $\mA$ is the algebra associated to a Rindler wedge, $\mA'$ is its commutant and $\frac{d}{dX^+}$ is the functional derivative with respect to the position of the Rindler wedge's entangling surface.

To prove this sum-rule without using density matrices, one needs to use the simple definition of the null momentum operator $P$ in terms of vacuum modular flow \eqref{eqn:differentiateP}, with the replacement $S_{\Omega} \to \Omega$ for a single region.
This can be taken as our abstract definition of the averaged null energy. One can then easily prove the sum-rule \eqref{eqn:normalsumrule} by using the ``differentiate'' definition of the relative entropy in terms of the cocycle
\begin{align}
S(\psi| \phi; \mA) = \lim_{t\to 0} \frac{1-\bra{\psi} (D\phi:D\psi;\mA)_t\ket{\psi}}{it}
\end{align}
together with the definition of $P$ in \eqref{eqn:differentiateP}. We leave this proof as an exercise for the reader.\footnote{It is important to note that these ``differentiate'' definitions of both the null energy and the relative entropy depend crucially on the assumption of finiteness of both quantities. For convenience, we will assume that all the relative entropies and null energies are finite. A proof of the sum rule \eqref{eqn:normalsumrule} that only assumes $\braket{P}_{\psi}<\infty$, $S(\psi | \Omega; \mA)<\infty$ and $S(\psi| \Omega; \mA')<\infty$ for $\mA$ but not necessarily any of its sub-algebras was given in Section 6 of \cite{CF}.}

To generalize equation \eqref{eqn:normalsumrule}, we can use the unitary $U:\mH \to \mH^{\otimes n}$ that maps between a single copy of the Hilbert space and $n$ copies. As discussed above, $\mA_r$ gets mapped to an algebra which acts on the $r$'th Hilbert space copy. The two states $\ket{\psi}$ and $\ket{\psi_s}$ then get mapped into an entangled state on two copies of the same QFT where the two states differ by the action of the cocycle in $\bigotimes_{r=1}^n \mA_r \equiv \mA$.

In the multi-copy Hilbert space, one can prove a formula exactly analogous to the sum-rule \eqref{eqn:normalsumrule} using the differentiate definition.\footnote{Actually, we really do need an analog of the more non-trivial proof of the sum-rule detailed in Section 6 of \cite{CF}. One can trivially map their proof to the multi-copy Hilbert space, however, and all the steps map accordingly. We leave this as an excercise for the reader.} For an arbitrary state $\phi \in \mH$, this new sum-rule takes the form
\begin{align}\label{eqn:doubledsumrule}
\left( \frac{d}{dX^+}\otimes \mathbf{1}\right)\left(S_{\text{rel}}\left(U\phi \,\left|\, \Omega^{\otimes n}\,;\mA'\right.\right) - S_{\text{rel}}\left(U\phi \,\left|\, \Omega^{\otimes n}\,;\mA\right.\right) \right) = \braket{P \otimes 1}_{U\phi}
\end{align}
where by $ \frac{d}{dX^+}\otimes \mathbf{1}$, we mean a functional derivative with respect to the position of the entangling surface associated to $A_1$ that has been mapped into the first copy, whose Hilbert space is $\mH \otimes \mathbf{1}$. 

If we now consider \eqref{eqn:doubledsumrule} in the state $\phi = \psi_s$, we can follow the steps laid out in Section 2 of \cite{CF} to reconstruct the relative entropy between $U\psi_s$ and $\Omega^{\otimes n}$ for general cuts of a Rindler wedge in each Hilbert space copy. In reconstructing the relative entropy, we will need the equation \eqref{eqn:splitnullenergy} for the split averaged null energy, discussed in the previous section \ref{sec:flowedANE}. What we find upon doing this is that in the multi-copy Hilbert space, the relative entropies $S(\psi_s|S_{\Omega};\mA)$ are differentiable with respect to their entangling surfaces' null positions.

Now, this differentiability does not obviously map back to differentiability in the single copy Hilbert space, at least in some neighborhood of the entangling surface. We can get around this issue, however, by using the trick from the previous subsection; we just pick our split state to be split with respect to a slightly large union of Rindler wedges. The algebras for this union of Rindler wedges is then $\tilde{\mA} = \bigcup_r\tilde{\mA}_r$ so that $\mA_r \subset \tilde{\mA}_r$. Differentiability for the slightly larger regions maps back to differentiability in a small neighborhood of the Rindler wedge $A_r$, which is all we need for formula \eqref{eq-energyresult} in the main text.

\section{Holographic derivation of the stress tensor shocks} \label{app:kink}

Here, we will provide a new derivation of the stress tensor shocks of Eq. \eqref{eqn:energyshocks} for holographic CFTs on Minkowski backgrounds in the special case of $n=2$. Note that here the bulk is merely providing an indirect tool for us to indirectly compute a boundary quantity. Let $a_{1}$ and $a_{2}$ to be two non-overlapping half-spaces in $d$ dimensional Minkowksi space. Now, consider a pure state $\psi$ dual to some asymptotically AdS$_{d+1}$ bulk geometry $\mathcal{M}$. We would like to transform the state by applying the unitary of Eq. \eqref{eq-flowbasic} which we reproduce here for the case of $n=2$:
\begin{align}\label{eq-apptrans}
    \ket{\psi_{s}}=\otimes_{r=1}^{2} \sigma^{is}_{a_{r}} (\rho^{\psi}_{a})^{-is} \ket{\psi}
\end{align}
where $\sigma_{a_{r}}$ denotes the vacuum density matrices of half-spaces $a_{r}$ and $a=a_{1} \cup a_{2}$.

It was shown in \cite{Bousso:2020yxi} that for a single half-space ($n=1$ case), the bulk dual of this transformation is given by the kink transform. We will now briefly describe the Kink transform following \cite{Bousso:2020yxi} and explain why we expect that it also provides the boundary dual of $\psi_s$ in Eq. \eqref{eq-apptrans}. We refer readers interested in the details of the Kink transform to \cite{Bousso:2020yxi}.
Let $\mathcal{R}$ be the RT surface of $a$ and let $\Sigma$ be a Cauchy slice of $\mathcal{M}$ containing $\mathcal{R}$. We can decompose $\Sigma$ in the following way:
\begin{align}
    \Sigma = \Sigma_{a} \cup \mathcal{R} \cup \Sigma_{\bar{a}}
\end{align}
such that $\mathcal{W}_{E}(a)=D(\Sigma_{a})$ and $\mathcal{W}_{E}(\bar{a})=D(\Sigma_{\bar{a}})$ where $\bar{a}$ is a complement of $a$ on the boundary.
The kink transform is defined as a particular change of the initial data $\Sigma \to \Sigma_{s}$ whose Cauchy evolution gives rise to a new asymptotically AdS$_{d+1}$ spacetime $\mathcal{M}_{s}$. Let $(h_{ab}, K_{ab})$ denote the initial data (the intrinsic metric and the extrinsic curvature respectively) on $\Sigma$. The kink transform yields new initial data  $\Sigma_{s}$ given by:
\begin{align}\label{eqn:kinktrans}
    \Sigma \to \Sigma_{s}: (h_{ab}, K_{ab} + x_{a} x_{b} \sinh(2\pi s) \delta(x))
\end{align}
where $x_{a}$ is the normal to $\mathcal{R}$ on $\Sigma$ and $x$ is the respective coordinate on $\Sigma_{s}$ in a neighborhood of $\mathcal{R}$. We also choose the initial data for other classical fields to be exactly the same on $\Sigma$ and $\Sigma_{s}$. Intuitively, there is a kink introduced at $\mathcal{R}$ which will cause a relative boost between the fields across $\mathcal{R}$. The Kink transform by definition preserves the initial data on $\Sigma_{a}$ and $\Sigma_{\bar{a}}$, so $\mathcal{M}_{s}$ contains the same $\mathcal{W}_{E}(a)$ and $\mathcal{W}_{E}(\bar{a})$ as in $\mathcal{M}$, but they are glued to each other with a relative ``boost'' at $\mathcal{R}$. This is why we expect that the kink transform around $\mathcal{R}$ describes the bulk dual of the boundary state $\psi_s$ since $\sigma_{a_r}^is$ generates such a relative boost in the $a$ region of the boundary with respect to its complement.

A a particular manifestation of the relative ``boost" caused by the Kink transform is the following. If one considers a smooth vector field $v^{\mu}$ on $\mathcal{R}$ with a smooth extension to the two entanglement wedges in $\mathcal{M}$, then in $\mathcal{M}_{s}$ the vector field would be discontinuous around $\mathcal{R}$. In particular, in \cite{Bousso:2020yxi} it was shown that the discontinuity can be represented as a local boost in the following sense:
\begin{align}
    v_{+}^{\mu} = (\Lambda_{2\pi s})^{\mu}_{\nu} v_{-}^{\nu}
\end{align}
where $v_{+}^{\mu}$ and $v_{-}^{\mu}$ corresponds to the limits of the vector field to $\mathcal{R}$ in $\mathcal{M}_{s}$ taken from the $\mathcal{W}_{E}(a)$ and $\mathcal{W}_{E}(\bar{a})$ sides respectively and $(\Lambda_{2\pi s})^{\mu}_{\nu}$ is a boost with rapidity $2\pi s$ acting in the normal bundle to $\mathcal{R}$.

We will now study the consequences of the kink transform in the near boundary region. In the Fefferman-Graham (FG) gauge, the metric in a neighborhood of the boundary entangling surfaces $\partial a_{r}$ takes the following form
\begin{align}\label{eq-FGmetric1}
    ds^{2} = \frac{1}{z_r^2} \left(dz_r^2 +\eta_{ij} dx_r^{i}dx_r^{j} + z_r^{d} \langle T_{ij}\rangle  + \mathcal{O}(z_r^{d+1}) \right)
\end{align}
where $i,j$ indices belong to constant $z$ slices and $\eta_{ij}$ is the Minkowski metric of the boundary:
\begin{align}
    ds_{\text{boundary}}^2 = -dt_r^2+dx_r^{2}+ \sum_{i=1}^{d-2} d{y_r^{i}}^2
\end{align}
where $t_{r}=a_{r}=0$ is $\partial a_{r}$ and $y_{r}$ parametrizes the boundary transverse (parallel to $\partial a_{r}$) direction. $a_{r}=\{x_r \geq 0\}$ in these coordinates. 
Now consider a bulk Cauchy slice $\Sigma$ containing $\mathcal{R}$ that anchors to the $t=0$ boundary slice. Let $t^{\mu}$ be the timelike unit normal to $\Sigma$ and $x^{\mu}$ be a unit vector field in the tangent space of $\Sigma$ at $\mathcal{R}$ and normal to it. The $z_r$ components of these vector fields in a neighborhood of $\partial a_{r}$ can be related to shape deformations of entropy in the boundary theory \cite{Koeller:2016aa, Akers:2017aa}. In particular
\begin{align}
    &x_{z_r}(y_r^i) = 4 G z_r^{d-2}  \left.\frac{\delta S(\rho^{\psi}_{a(X_r)})}{\delta X_r(y^i_r)} \right|_{X_r=0} + o(z_r^{d-2}),\\
    &t_{z_r}(y_r^i) = 4 G z_r^{d-2}  \left.\frac{\delta S(\rho^{\psi}_{a(T_r)})}{\delta T_r(y^i_r)} \right|_{T_r=0} + o(z_r^{d-2}).
\end{align}
where $a(X_{r})$ and $a(T_{r})$, in analogy with Sec. \ref{sec:cc}, is a deformation of $a$ by moving $\partial a_{r}$ to $\{x_r = X(y_r^i), t_r=0\}$ and $\{x_r = 0, t_r = T(y_r^i)\}$ respectively. $G$ is the bulk Newton's constant.

In $\mathcal{M}_{s}$, we will insist on using the same FG coordinates as above in $\mathcal{W}_{E}(a)$ and $\mathcal{W}_{E}(\bar{a})$ in a bulk neighborhood of $\partial a_r$,\footnote{We get to choose the same coordinates because the Kink transform leaves the geometries unchanged in $\mathcal{W}_{E}(a)$ and $\mathcal{W}_{E}(\bar{a})$.} but since this is a new spacetime we re-label these coordinates as ($\tilde{t}_r, \tilde{x}_r, \tilde{y}_r^{i}, \tilde{z}_r)$. These tilde coordinates also extend to the boundary in which the kinked boundary anchor of $\Sigma_{s}$ is located at $\tilde{t}_r=\tilde{z}_r=0$. The boundary metric in the tilde coordinates is:
\begin{align}\label{eq-crazymetric}
    ds_{\text{boundary}}^{2} &= [\Theta(\tilde{t}_r+\tilde{x}_r) + e^{2\pi s}(1-\Theta(\tilde{t}_r+\tilde{x}_r)][e^{-2\pi s} \Theta(\tilde{t}_r-\tilde{x}_r)+(1-\Theta(\tilde{t}_r-\tilde{x}_r))](-d\tilde{t}_r^2+d\tilde{x}_r^2)\nonumber\\
    &+ \sum_{i=1}^{d-2} {d\tilde{y}_r^i}^2.
\end{align}

We can simply check using these coordinates that the $\tilde{t}_r=0$ slice, even though it is smoothly specified in the tilde coordinates, has an extrinsic curvature shock as demanded by the kink transform. This is due to Christoffel symbol shocks in the metric \eqref{eq-crazymetric}. For example,
\begin{align}\label{eq-christoffels1}
    \Gamma^{\tilde{t}_r}_{\tilde{x}_r\tilde{x}_r}=\sinh (2\pi s) \delta(\tilde{x}_r)
\end{align}
The bulk kink at $\mathcal{R}$ looks locally the same as that of the boundary anchor, therefore in the tilde FG coordinates there must exists analogous Christoffel shocks around $\mathcal{R}$. In particular,\footnote{The bulk Christoffel shocks in Eq. \eqref{eq-christoffels2} are stronger conditions than the bulk version of Eq. \eqref{eq-christoffels1}, but we will leave their derivation as an exercise. Furthermore, we expect that these Christoffel shocks can also be directly derived from the definition of the kink transform in \eqref{eqn:kinktrans}.}
\begin{align}\label{eq-christoffels2}
    &\tilde{t}_{\mu} \tilde{x}^{\nu} \Gamma^{\mu}_{\nu \alpha} = \sinh (2\pi s) \tilde{x}_{\alpha} \delta(\tilde{x})+ o(\delta),\\
    &\tilde{x}^{\mu} \tilde{x}^{\nu} \Gamma^{\alpha}_{\mu \nu} = \sinh (2\pi s) \tilde{t}^{\alpha} \delta(\tilde{x}) + o(\delta).
\end{align}
where $\tilde{x}^{\alpha}$ is the unit normal vector to $\mathcal{R}$ in the tilde coordinates. Taking $\alpha=z$ and using \eqref{eq-FGmetric1}, we can relate these Christoffel shocks to CFT stress tensor shocks. We find at the entangling surface
\begin{align}\label{eqn:shocktilde}
   &\langle T_{\tilde{t}\tilde{x}}(\tilde{t}_r=0, \tilde{x}_r, y^i_r)\rangle = \frac{1}{2\pi} \sinh (2\pi s) \left.\frac{\delta S(\rho^{\psi}_{a(\tilde{X}_r)})}{\delta \tilde{X}_r(y^i_r)} \right|_{\tilde{X}_r=0}\delta (\tilde{x}_r)+ o(\delta),\\ 
    &\langle T_{\tilde{x}\tilde{x}}(\tilde{t}_r=0, \tilde{x}_r, y^i_r)\rangle= \frac{1}{2\pi} \sinh (2\pi s) \left.\frac{\delta S(\rho^{\psi}_{a(\tilde{T}_r)})}{\delta \tilde{T}_r(y^i_r)} \right|_{\tilde{T}_r=0}\delta (\tilde{x}_r)+ o(\delta).
\end{align}
In order to compare the results to Sec. \ref{sec:cc}, we can now change boundary coordinates from $(\tilde{t}_r, \tilde{x}_r, \tilde{y}^{i}_r)$ to coordinates $(x_r^{-},x_r^{+},y_r^{i})$ in which $a_r=\{x_r^+ -x_r^- \geq 0\}$ and $ ds_{\text{boundary}}^2=-dx_r^+ dx_r^- +\sum_{i=1}^{d-2} (dy^i_r)^2$ to compare the results to that of Sec. \ref{sec:cc}:
\begin{align}
    &x_r^{+}=(\tilde{t}_r+\tilde{x}_r)\Theta(\tilde{t}_r+\tilde{x}_r)+e^{-2\pi s}(\tilde{t}_r+\tilde{x}_r)(1-\Theta(\tilde{t}_r+\tilde{x}_r))\nonumber \\
    &x_r^{-}=e^{2\pi s}(\tilde{t}_r+\tilde{x}_r)\Theta(\tilde{t}_r-\tilde{x}_r)+(\tilde{t}_r-\tilde{x}_r)(1-\Theta(\tilde{t}_r-\tilde{x}_r))\\
    &y_r^{i} = \tilde{y}_r^{i}.
\end{align}
Changing coordinates in \eqref{eqn:shocktilde} gives the following shocks at $\partial a$ on the boundary
\begin{align}
    &\langle T_{++}(x^+_r, x^-_r=0, y^i_r) \rangle= \frac{1}{2\pi}(e^{-2\pi s}-1) \left.\frac{\delta S(\rho^{\psi}_{a(X^{+}_r)})}{\delta X^+_r(y^{i})}\right|_{X^+_r=0} \delta(x^{+}_r)+o(\delta)\nonumber \\
    &\langle T_{--}(x^+_r=0, x^-_r, y^i_r)\rangle = \frac{1}{2\pi}(e^{2\pi s}-1) \left.\frac{\delta S(\rho^{\psi}_{a(X^{-}_r)})}{\delta X^-_r(y^{i})}\right|_{X^-_r=0} \delta(x^{-}_r)+o(\delta)\nonumber \\
    &\langle T_{+-}(x^+_r, x^-_r, y^i_r)\rangle = o(\delta) .
\end{align}
This is consistent with the results in Sec. \ref{sec:cc}.

\section{JT Charges Review} \label{app:jt}
%%%%%%%%%%%%%%%%%%%%%%%%%%%%%%%%%%%%%%%%
JT gravity is a solvable theory of 2d gravity that manifests some important properties of AdS/CFT.
In Lorentzian signature, it has action
\begin{equation}
  S_{JT} = \frac{\phi_0}{4\pi} \left\{ \int_{M} \sqrt{-g} R + \int_{\partial M} \sqrt{-h} K \right\} + \frac{1}{4\pi} \left\{ \int_{M} \sqrt{-g}\phi (R+2) - 2\int_{\partial M} \sqrt{-h} \phi (K-1) \right\},
  \label{eqn:jt-action}
\end{equation}
with boundary conditions
\begin{equation}
  ds^{2}|_{\partial M} = - \frac{du^{2}}{\epsilon^{2}}, \quad \phi|_{\partial M} = \frac{\phi_{r}}{\epsilon}.
  \label{eqn:jt-bd-conds}
\end{equation}
The first condition should be thought of as a definition of the physical time $u$.
The $\phi$ equation of motion sets the geometry to be locally AdS${}_{2}$ everywhere, and so all the dynamics is that of the dilaton; because of the boundary condition, this can further be reduced to that of the trajectory of the boundary.
The rest of this appendix is devoted to a short review of the elegant $SL(2,\mathbb{R})$ charge formalism for the dynamics of this boundary particle \cite{MSY-1,MSY-2,LMZ}.

In a 3d embedding space with metric
\begin{equation}
  ds_{3}^{2} = dY_{a} dY^{a} = - \left( dY^{-1} \right)^{2} - \left( dY^{0} \right)^{2} + \left( dY^{1} \right)^{2}
  \label{eqn:embedding-metric}
\end{equation}
AdS${}_{2}$ is the 2d surface
\begin{equation}
  - \left( Y^{-1} \right)^{2} - \left( Y^{0} \right)^{2} + \left( Y^{1} \right)^{2} = -1.
  \label{eqn:ads2-surface}
\end{equation}
The three isometries of AdS${}_{2}$ are merely the generators of the 3d Lorentz group.
The relation between these and the Kruskal-Szekeres coordinates used in the main text is
\begin{equation}
  (Y^{-1},Y^{0},Y^{1}) = \left( \frac{1 - w^{+} w^{-}}{1 + w^{+} w^{-}} , \frac{w^{+} + w^{-}}{1 + w^{+} w^{-}}, \frac{w^{+} - w^{-}}{1 + w^{+} w^{-}} \right).
  \label{eqn:kruskal-global-reln}
\end{equation}
The three Killing vectors are
\begin{align}
  \zeta^{-1} &= w^{+} \partial_{+} - w^{-} \partial_{-} \nonumber\\
  \zeta^{+} = \zeta^{0} + \zeta^{1} &= (w^{+})^{2} \partial_{+} + \partial_{-} \nonumber\\
  \zeta^{-} = \zeta^{0} - \zeta^{1} &= - \partial_{+} - (w^{-})^{2} \partial_{-}.
  \label{eqn:KVs}
\end{align}

In embedding coordinates, the general solution for the dilaton with no stress tensor is
\begin{equation}
  \phi = - Q \cdot Y.
  \label{eqn:dilaton-soln}
\end{equation}
Because of the boundary condition \eqref{eqn:jt-bd-conds}, we then have for the boundary coordinates
\begin{equation}
  Q \cdot X = -  \phi_r, \quad X \equiv \lim_{\epsilon \to 0} \epsilon Y,\quad Q^{2} = - 4 \phi_r M = - \left( \frac{2\pi \phi_r}{\beta} \right)^2.
  \label{eqn:bd-particle-trajectory}
\end{equation}
Here, $M$ is the AdM energy of the bulk, and $\beta$ is the corresponding inverse temperature; the latter expression is useful since $\beta$ doesn't scale with $\phi_r$.
This charge also locates the horizon,
\begin{equation}
  Y_{l}^{a} = \frac{Q_{l}^{a}}{\sqrt{- Q^{2}}}, \quad Y_r^a =  -\frac{Q_r^a}{\sqrt{-Q^2}},
  \label{eqn:horizon-loc}
\end{equation}
where we've chosen the convention that $-Q_r^{1} \ge 0$.

The backreaction of a matter stress tensor can be thought of as a spatially varying charge.
Since we're interested in only the boundary particles, it is enough to know that
\begin{equation}
  Q_{l}^{a} + Q_{r}^{a} +  Q_{mat}^{a} = 0.
  \label{eqn:gauge-cond}
\end{equation}
and
\begin{equation}
  Q_{mat}^{a} = 2\pi \int d\Sigma^{\mu} (\zeta^{a})^{\nu} T_{\mu\nu}.
  \label{eqn:matter-charge}
\end{equation}
Another way to determine the charge of a point particle of mass $m$ with trajectory $Y^{a} (s)$ is
\begin{equation}
  Q_{a} = \varepsilon_{abc} Y^{b} d_{s} Y^{c}, \quad \left( d_{s} Y \right)^{2} = - m^{2}.
  \label{eqn:matter-charge-pt}
\end{equation}

Finally, we list charges of some configurations that are useful for us.
For an eternal black hole of inverse temperature $\beta$ with no bulk stress tensor,
\begin{equation}
  Q_{l}^{a} = - Q_{r}^{a} = \frac{2\pi \phi_r}{\beta} (1,0,0).
  \label{eqn:tfd-charge}
\end{equation}
For a massless particle that passes through $Y^{0} = 0, Y^{1} = y$ with velocity $\pm 1$, the charge is
\begin{equation}
  q_{\pm}^{a} =\frac{p_{\mp}}{2} \left( y , \pm 1, \sqrt{1 + y^{2}} \right), \quad p_{\pm} = 2\pi T_{w^{\pm} w^{\pm}}.
  \label{eqn:shock-charge}
\end{equation}

%%%%%%%%%%%%%%%%%%%%%%%%%%%%%%%%%%%%%%%%

\bibliographystyle{JHEP}
\bibliography{all}

\end{document}